\documentclass[11pt]{article} 
\usepackage{amsmath,amssymb} 
\usepackage[dvips]{graphicx,color}

\numberwithin{equation}{section}

\begin{document}

\title{A Mathematical Analysis of Dressed Photon in Ground State \\ 
of Generalized Quantum Rabi Model Using Pair Theory}

\author{Masao Hirokawa \\ 
{\footnotesize Institute of Engineering, Hiroshima University, Higashi-Hiroshima, 739-8527, Japan} \\ 
Jacob S. M{\o}ller \\ 
{\footnotesize Department of Mathematics, Aarhus University, Aarhus, Denmark} \\ 
Itaru Sasaki \\ 
{\footnotesize Department of Mathematical Sciences, Shinshu University, Matsumoto, 390-8621, Japan}
}

\date{}
\maketitle

\abstract{
We consider the generalized quantum Rabi model with the 
so-called $A^{2}$-term in the light of the Hepp-Lieb-Preparata 
quantum phase transition. 
We investigate the dressed photon in its ground state 
when the atom-light coupling strength is 
in the deep-strong coupling regime. 
We show how the dressed photon appears in the ground state. 
We dedicate this paper to Pavel Exner and Herbert Spohn 
on the occasion of their 70th birthdays, and Klaus Hepp 
on the occasion of his 80th birthday. 
}

\section{Introduction}
\label{sec:intro}

Quantum electrodynamics (QED) says that the matter coupled with 
a radiation field emits a photon when the relaxation (de-excitation) 
of a quantum state of the matter takes place. 
In particular, for the relaxation from an excited state 
to a ground state, the ground state should be the vacuum. 
Preparata, however, claims that there is a coherence domain 
in which the photon cannot be emitted from the matter 
in certain cases where the matter-radiation coupling 
is very strong \cite{preparata,enz}, and the ground state 
has the photon that should primarily be emitted 
outside of the matter. 
Thus, the photon should be among the dressed photons in the ground state. 
Then, the ground state switch from the perturbative ground state 
to the non-perturbative ground state. 
This non-perturbative ground state is called the coherent ground state 
by Preparata \cite{preparata} and the superradiant ground state by Enz \cite{enz}. 
Preparata found this phenomena stimulated by and based on 
the Hepp-Lieb quantum phase transition \cite{hepp-lieb1,hepp-lieb2}. 
We thus call this phenomenon the Hepp-Lieb-Preparata quantum phase transition 
in this paper. 

The Hepp-Lieb-Preparata quantum phase transition is shown by Enz \cite{enz} 
and one of authors \cite{hiro-rmp} 
for a model describing a two-level system coupled with light 
such as the many-mode-photon version of the Jaynes-Cummings type models 
as well as Jaynes-Cummings model itself \cite{hiro-pla,hiro-pra,hiro-iumj}.  
We note that all the models have the rotating wave approximation. 
However, in the ultra-strong or deep-strong 
coupling regime \cite{casanoba-solano} 
we cannot avoid the effects coming from 
the counter rotating terms and the $A^{2}$-term, 
the quadratic interaction of photon field, 
for the argument on such a quantum phase transition. 
It is worthy to note that some theoretical observations 
have lately pointed out the possibility of the quantum phase transition 
in a ground state of a model in circuit QED \cite{nori,ciuti,ciuti2}.

For the model with the $A^{2}$-term and without the rotating wave approximation, 
Yoshihara \textit{et al}. recently succeeded in achieving 
the deep-strong coupling regimes in their circuit QED experiment. 
They observed a quantum phase transition in the ground state 
in the deep-strong coupling regime for the generalized quantum Rabi model. 
Therefore, we are mathematically interested in the generalized quantum Rabi model, 
and we consider it in the light of the Hepp-Lieb-Preparata quantum phase transition.  
There are at least two possibilities: whether the dressed photon in the ground state 
is a bare photon or it is a physical photon.  
Because, in particular, we pay special attention to the physical photon in the ground state, 
we cope with the $A^{2}$-term problem 
following the pair theory for quantum field caused by 
a neutral source \cite{henley-thirring}. 
For the history of several pair theories, see Ref.\cite{br}.

\section{Our Model and Problems} 

\subsection{Notations}
\label{sec:NN}

Let $\mathfrak{H}$ be a separable Hilbert space. 
We denote by $(\,\,\, , \,\,\,)_{\mathfrak{H}}$ the inner product of 
the Hilbert space $\mathfrak{H}$.  
We denote by $\|\,\,\, \|_{\mathfrak{H}}$ the norm 
naturally induced by the inner product $(\,\,\, , \,\,\,)_{\mathfrak{H}}$. 
Meanwhile, $\|\,\,\,\|_{\mathrm{op}}$ is the operator norm 
for a bounded operator acting the Hilbert space. 
The state space $\mathcal{F}$ of the two-level atom system coupled 
with one-mode light is given by $\mathbb{C}^{2}\otimes L^{2}(\mathbb{R})$, 
where $\mathbb{C}^{2}$ is the $2$-dimensional unitary space, 
and $L^{2}(\mathbb{R})$ the Hilbert space consisting of 
the square-integrable functions. 
For any matrix $A$ on $\mathbb{C}^{2}$ and 
any operator $B$ acting in $L^{2}(\mathbb{R})$, 
we often denote by $AB$ the tensor product of the operators 
$A$ and $B$, i.e., $AB=A\otimes B$, 
by omitting the tensor sign $\otimes$ throughout this paper.  
For the identity operator $I$, we write $A\otimes I$ and $I\otimes B$ 
simply as $A$ an $B$, respectively.  
We use the notations, $|\!\!\uparrow\rangle$ and $|\!\!\downarrow\rangle$,  
which stand for the spin-states defined in $\mathbb{C}^{2}$ by 
$
|\!\!\uparrow\rangle
:=
{\scriptsize 
\left(\hspace*{-1.5mm}
\begin{array}{cc}
1 \\ 
0 
\end{array}
\hspace*{-1.5mm}
\right)
}$
and 
$ 
|\!\!\downarrow\rangle
:=
{\scriptsize 
\left(\hspace*{-1.5mm}
\begin{array}{cc}
0 \\ 
1 
\end{array}
\hspace*{-1.5mm}
\right)
}$. 
We denote by $|n\rangle$ the Fock state in $L^{2}(\mathbb{R})$ 
with the photon number 
$n=0, 1, 2, \cdots$. 
That is, $|0\rangle:=(\omega/\pi\hbar)^{1/4}
\exp\left[-\omega x^{2}/2\hbar\right]$ and 
$|n\rangle:=\sqrt{w}\gamma_{n}H_{n}(wx)\exp\left[-(wx)^{2}/2\right] 
\in L^{2}(\mathbb{R})$, 
where  $H_{n}(x)$ is the Hermite polynomial of variable $x$, 
$\gamma_{n}=\pi^{-1/4}(2^{n}n!)^{-1/2}$, and 
$w=\sqrt{m\omega_{\mathrm{c}}/\hbar}$.
We often use a compact notation, $|s,n\rangle$, 
for the separable state $|s\rangle\otimes |n\rangle$ 
for $s=\uparrow, \downarrow$ and $n=0, 1, 2, \cdots$. 

We denote the Pauli matrices by 
$\sigma_{x}:=
{\scriptsize 
\left(\hspace*{-1.5mm}
\begin{array}{cc}
0 & 1 \\ 
1 & 0
\end{array}
\hspace*{-1.5mm}\right)
}$, 
$\sigma_{y}:=
{\scriptsize 
\left(\hspace*{-1.5mm}
\begin{array}{cc}
0 & -i \\ 
i & 0
\end{array}
\hspace*{-1.5mm}\right)
}$, 
and  
$\sigma_{z}:=
{\scriptsize 
\left(\hspace*{-1.5mm}
\begin{array}{cc}
1 & 0 \\ 
0 & -1
\end{array}
\hspace*{-1.5mm}
\right)
}$. 
We respectively denote by $a$ and $a^{\dagger}$ 
the annihilation and creation operators of $1$-mode photon 
defined by $a|0\rangle:=0$, $a|n\rangle:=\sqrt{n}|n-1\rangle$, 
and $a^{\dagger}|n\rangle:=\sqrt{n+1}|n+1\rangle$. 

We denote by $|E\rangle$ a normalized eigenstate of an operator $A$ with 
its corresponding eigenvalue $E$, i.e., $A|E\rangle=E|E\rangle$. 
For any state $\psi$, 
we denote $\sigma_{x}\psi$ by $\widetilde{\psi}$, i.e., 
$\widetilde{\psi}=\sigma_{x}\psi$. 
For any eigenstate $|E\rangle$, 
we denote  $\sigma_{x}|E\rangle$ by $|\widetilde{E}\rangle$. 
We note that the operator $\widetilde{A}$ defined by $\widetilde{A}
=\sigma_{x}A\sigma_{x}$ is unitarily equivalent to the operator $A$ 
so that $|\widetilde{E}\rangle$ is its eigenstate with the corresponding 
eigenvalue $E$.

\subsection{Mathematical Models}

The flux qubit is demonstrated in 
a superconducting circuit. 
The clockwise current and the counterclockwise current in 
the superconducting circuit make two states, 
$|\!\!\circlearrowright\rangle$ and $|\!\!\circlearrowleft\rangle$, respectively. 
We can regard these two states as a qubit. 
Then, the Hamiltonian of the qubit is given by 
$$
\mathcal{H}_{\mathrm{qubit}}=\, -\, \frac{\hbar}{2}\left(\omega_{\mathrm{a}}\sigma_{x}+\varepsilon\sigma_{z}\right). 
$$
%%% \begin{center}
%%% \begin{figure}
%%% % Use the relevant command to insert your figure file.
%%% % For example, with the graphicx package use
%%% \begin{center}
%%%   \includegraphics[width=0.4\textwidth]{flux_qubit.eps}
%%% \end{center}
%%% % figure caption is below the figure
%%% %\vspace*{12mm}
%%% \caption{Flux qubit. A Josephson junction is used instead of the condenser 
%%% in a superconducting LC circuit. 
%%% Applying a magnetic field so that it passes outside or inside the circuit, 
%%% the current in the circuit becomes clockwise or counterclockwise. }
%%% \label{fig:flux_qubit}       % Give a unique label
%%% \end{figure}
%%% \end{center}
Here, $\hbar\omega_{\mathrm{a}}$ and $\hbar\varepsilon$ 
are the tunneling splitting energy and the bias energy between 
$|\!\!\circlearrowright\rangle$ and $|\!\!\circlearrowleft\rangle$. 
We respectively represent the states $|\!\!\circlearrowright\rangle$ and 
$|\!\!\circlearrowleft\rangle$ with the spin-states  $|\!\!\downarrow\rangle$ and 
$|\!\!\uparrow\rangle$ from now on: 
$|\!\!\downarrow\rangle=|\!\!\circlearrowright\rangle$ and 
$|\!\!\uparrow\rangle=|\!\!\circlearrowleft\rangle$.  
Then, we can regard the Pauli matrix $\sigma_{x}$ as the \textit{spin-chiral transformation}, 
$\sigma_{x}:|\!\!\downarrow\rangle=|\!\!\circlearrowright\rangle 
\longrightarrow |\!\!\uparrow\rangle=|\!\!\circlearrowleft\rangle$ and 
$\sigma_{x}:|\!\!\uparrow\rangle=|\!\!\circlearrowleft\rangle 
\longrightarrow |\!\!\downarrow\rangle=|\!\!\circlearrowright\rangle$.  
%%% $$
%%% \sigma_{x} : 
%%% \begin{array}{l}
%%% |\!\!\downarrow\rangle=|\!\!\circlearrowright\rangle 
%%% \longrightarrow |\!\!\uparrow\rangle=|\!\!\circlearrowleft\rangle, \\ 
%%% |\!\!\uparrow\rangle=|\!\!\circlearrowleft\rangle 
%%% \longrightarrow 
%%% |\!\!\downarrow\rangle=|\!\!\circlearrowright\rangle. 
%%% \end{array}
%%% $$
In other words, for the counterclockwise rotation matrix $R_{z}(\theta)=
e^{-i\theta\sigma_{z}/2}$ and  the clockwise rotation matrix 
$R_{z}(-\theta)=e^{i\theta\sigma_{z}/2}$ through an angle $\theta$ about the $z$-axis 
of the Bloch sphere \cite{BCS}, we have the relation, 
$\sigma_{x}R_{z}(\theta)\sigma_{x}=R_{z}(-\theta)$, 
and therefore, we can say that the states, $\widetilde{\psi}$ and 
$|\widetilde{E}\rangle$, are the reflected images 
on the other side of mirror on  the $xy$-surface 
for their original states, 
$\psi$ and $|E\rangle$, respectively.

Yoshihara \textit{et al}. consider the Hamiltonian
$$
\mathcal{H}_{\mathrm{total}}
=\mathcal{H}_{\mathrm{qubit}}
+\hbar\omega_{\mathrm{c}}\left(a^{\dagger}a+\frac{1}{2}\right)
+\hbar\mathrm{g}\sigma_{z}\left(a+a^{\dagger}\right) 
$$
in Ref.\cite{semba2016}. 
Let us define a unitary operator $U_{xz}$ by   
$
U_{xz}:=\frac{1}{\sqrt{2}}
{\scriptsize 
\left(\hspace*{-1.5mm}
\begin{array}{cc}
1 & 1 \\ 
-1 & 1
\end{array}
\hspace*{-1.5mm}\right)
}$. 
Then, their Hamiltonian $\mathcal{H}_{\mathrm{total}}$ is unitarily transformed to 
\begin{equation}
H(\omega_{\mathrm{a}},\varepsilon,\omega_{\mathrm{c}}, \mathrm{g})
:=
U_{xz}^{*}\mathcal{H}_{\mathrm{total}}U_{xz} 
=
 H_{\mathrm{atom}}(\omega_{\mathrm{a}},\varepsilon)
+H_{\mathrm{cavity}}(\omega_{\mathrm{c}})+H_{\mathrm{int}}(\mathrm{g})
\label{eq:total-Hamiltonian}
\end{equation}
with 
$H_{\mathrm{atom}}(\omega_{\mathrm{a}},\varepsilon)=
\frac{\hbar}{2}\left(\omega_{\mathrm{a}}\sigma_{z}-\varepsilon\sigma_{x}\right)$,  
$H_{\mathrm{cavity}}(\omega_{\mathrm{c}})=
\hbar\omega_{\mathrm{c}}\left(a^{\dagger}a+\frac{1}{2}\right)$, 
and 
$H_{\mathrm{int}}(\mathrm{g})=
\hbar\mathrm{g}\sigma_{x}\left(a+a^{\dagger}\right)$. 
%%% $$
%%% \cases{
%%% H_{\mathrm{atom}}(\omega_{\mathrm{a}},\varepsilon)=
%%% \frac{\hbar}{2}\left(\omega_{\mathrm{a}}\sigma_{z}-\varepsilon\sigma_{x}\right), \\
%%% H_{\mathrm{cavity}}(\omega_{\mathrm{c}})=
%%% \hbar\omega_{\mathrm{c}}\left(a^{\dagger}a+\frac{1}{2}\right), \\ 
%%% H_{\mathrm{int}}(\mathrm{g})=
%%% \hbar\mathrm{g}\sigma_{x}\left(a+a^{\dagger}\right).   
%%% }
%%% $$
Here the cavity Hamiltonian $H_{\mathrm{cavity}}(\omega_{\mathrm{c}})$ 
includes the zero-point energy $\hbar\omega_{\mathrm{c}}/2$.
In the case $\varepsilon=0$, we call it the \textit{quantum Rabi Hamiltonian}, 
and thus, we call $H(\omega_{\mathrm{a}},\varepsilon,\omega_{\mathrm{c}},\mathrm{g})$ 
a \textit{generalized quantum Rabi Hamiltonian}. 
We rewrite the interaction in the (generalized) quantum Rabi Hamiltonian 
using the spin-annihilation operator $\sigma_{-}$ 
and the spin-creation operator $\sigma_{+}$ 
defined by $\sigma_{\pm}
:=(\sigma_{x}\pm i\sigma_{y})/2$: 
$\hbar\mathrm{g}\sigma_{x}\left(a+a^{\dagger}\right)
=\hbar\mathrm{g}
\left(
\sigma_{-}a+\sigma_{+}a+\sigma_{-}a^{\dagger}+\sigma_{+}a^{\dagger}
\right)$. 
The operators, $\sigma_{+}a$ and $\sigma_{-}a^{\dagger}$, are the rotating terms, 
and the operators, $\sigma_{-}a$ and $\sigma_{+}a^{\dagger}$, the counter-rotating terms. 
The spin-chiral transformation $\sigma_{x}$ makes the relations, 
$\sigma_{x}(\sigma_{-}a)\sigma_{x}=\sigma_{+}a$ 
and 
$\sigma_{x}(\sigma_{+}a^{\dagger})\sigma_{x}=\sigma_{-}a^{\dagger}$. 
Thus, the counter-rotating terms are the rotating terms
on the other side of the mirror.
%%%  as in Fig.\ref{fig:sct}. 
%%% \begin{center}
%%% \begin{figure}
%%% \begin{center}
%%%   \includegraphics[width=0.2\textwidth]{mirror.eps}
%%% \end{center}
%%% \caption{Spin-chiral transformation of counter-rotating terms.}
%%% \label{fig:sct}       % Give a unique label
%%% \end{figure}
%%% \end{center}

We here give some mathematical notes. 
By Theorem 4.3 of Ref.\cite{hiro-iumj}, 
the generalized quantum Rabi Hamiltonian 
$H(\omega_{\mathrm{a}},\varepsilon,\omega_{\mathrm{c}},\mathrm{g})$ 
is self-adjoint on the domain $\mathbb{C}^{2}\otimes 
\mathrm{Span}(\left\{|n\rangle\right\}_{n=0}^{\infty})$, 
where $\mathrm{Span}(\left\{|n\rangle\right\}_{n=0}^{\infty})$ is 
a subspace linearly spanned by all the Fock states $|n\rangle$ in $L^{2}(\mathbb{R})$.

\subsection{Hepp-Lieb-Preparata Quantum Phase Transition}
\label{sec:hlp}

We briefly see the Hepp-Lieb-Preparata quantum phase transition in mathematics 
using the Jaynes-Cummings model though the model is not physically valid over the 
ultra-strong coupling regime. 
For simplicity we tune the frequencies, $\omega_{\mathrm{a}}$ and $\omega_{\mathrm{c}}$, 
to a frequency $\omega$, i.e., $\omega_{\mathrm{a}}=\omega_{\mathrm{c}}=\omega$. 
The Jaynes-Cummings Hamiltonian reads 
$$
H_{\mathrm{JC}}:=
H_{\mathrm{atom}}(\omega,0)+H_{\mathrm{cavity}}(\omega)+
\hbar\mathrm{g}(\sigma_{+}a+\sigma_{-}a^{\dagger}). 
$$
It is easy to solve the eigenvalue problem, $H_{\mathrm{JC}}\varphi_{\nu}=E_{\nu}\varphi_{\nu}$. 
The eigenstates $\varphi_{\nu}$ and the corresponding eigenvalues $E_{\nu}$ are given by 
$$
\left\{
\begin{array}{ll}
{\displaystyle 
\varphi_{0}= 
|\!\!\downarrow, 0\rangle}, \\ 
{\displaystyle 
E_{0}=0},
\end{array}
\right.
\quad
\textrm{and}
\quad 
\left\{
\begin{array}{ll}
{\displaystyle 
\varphi_{\pm (n+1)}=
\pm\frac{1}{\sqrt{2}}|\!\!\uparrow, n\rangle 
+\frac{1}{\sqrt{2}}|\!\!\downarrow, n+1\rangle},   \\ 
{\displaystyle 
E_{\pm (n+1)}= 
\hbar\omega (n+1)\pm\hbar\mathrm{g}\sqrt{n+1}}, 
\end{array}
\right. 
\quad 
n=0, 1, 2, \cdots. 
$$
Thus, we realize that the separable state 
$\varphi_{0}$ is a unique ground state if $0\le \mathrm{g}<\omega$, 
and the entangled state $\varphi_{-1}$ is a unique ground state 
if $\omega<\mathrm{g}<(\sqrt{2}+1)\omega$. 
In the same way, moreover, 
the entangled state $\varphi_{-(n+1)}$ is a unique ground state 
if $(\sqrt{n+1}+\sqrt{n})\omega<\mathrm{g}<
(\sqrt{n+2}+\sqrt{n+1})\omega$. 
In these way, the ground state has some photons provided 
$\omega<\mathrm{g}$, 
though it has no photon as long as $0\le \mathrm{g}<\omega$. 
The ground state shifts from a separable state to an entangled one, 
and changes how to make the entanglement.  
Namely, we can concretely see the Hepp-Lieb-Preparata quantum phase transition 
for the Jaynes-Cummings model
through the facts how many photons the ground state has, 
and how the entanglement forms the non-perturbative ground state. 

We are interested in the quantum Rabi model's case. 
We now tune the frequencies, $\omega_{\mathrm{a}}$ and $\omega_{\mathrm{c}}$, 
as in the above. 
As shown in Ref.\cite{hiro-qs}, the free part of the quantum Rabi Hamiltonian $H(\omega,0,\omega,0)$ 
is the Witten Laplacian \cite{witten} which describes the $N=2$ SUSY quantum mechanics, 
and the renormalized Rabi Hamiltonian 
$H(\omega,0,\omega,\mathrm{g})+\hbar\mathrm{g}^{2}/\omega$ 
converges to the Hamiltonian describing the spontaneously $N=2$ SUSY breaking 
in the weak operator topology. 
Here, the energy $-\hbar\mathrm{g}^{2}/\omega$ is 
the binding energy of the van Hove Hamiltonians 
$H_{\mathrm{cavity}}(\omega)\pm\hbar\mathrm{g}(a+a^{\dagger})$ 
appearing in neutral static-source theory. 
Actually, the individual van Hove Hamiltonian sits 
in each diagonal element of the total Hamiltonian 
$\mathcal{H}_{\mathrm{total}}\lceil_{\varepsilon=0}$; 
it is referred to as an energy renormalization 
(see Ch.9 of Ref.\cite{henley-thirring}).   
The renormalized Rabi Hamiltonian recovers the spin-chiral symmetry 
as the coupling strength grows lager and larger, and then, the spin-chirality causes 
spontaneously SUSY breaking. 
For more details, see Ref.\cite{hiro-qs}. 

\begin{center}
\begin{figure*}
\begin{center}
\includegraphics[width=0.3\textwidth]{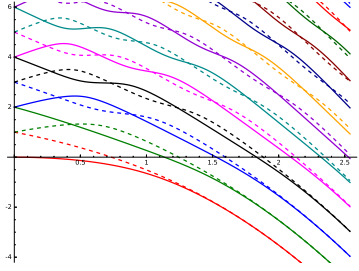}
\vspace*{12mm}
\caption{
The energy revel of the quantum Rabi Hamiltonian 
$H(\omega, 0, \omega, \mathrm{g})$. 
The unit of the transverse axis is $\mathrm{g}/\omega$, 
and the unit of the longitudinal axis is $\hbar\omega$.
} 
\label{fig:energy_rabi}       % Give a unique label
\end{center}
\end{figure*}
\end{center}

Here, we improve the quantum Rabi Hamiltonians' convergence; 
it is actually in the norm resolvent sense 
(see Definition on p.284 of Ref.\cite{rs1}).  
That is, as proved below, we can obtain the following convergence: 
\begin{equation}
\lim_{\mathrm{g}\to\infty}
\| 
\left(H\left(\omega_{\mathrm{a}},0,\omega_{\mathrm{c}},\mathrm{g}\right)
+\hbar\mathrm{g}^{2}/\omega_{\mathrm{c}}
-i\right)^{-1}
-
\left( U_{\mathrm{g}}^{*}
H_{\mathrm{cavity}}(\omega_{\mathrm{c}})
U_{\mathrm{g}}-i
\right)^{-1}
\|_{\mathrm{op}} 
=0, 
\label{eq:norm-resolvent-cov}
\end{equation}
where $U_{\mathrm{g}}$ 
is the Bogoliubov transformation 
defined by the displacement operator, 
$U_{\mathrm{g}}:=e^{\mathrm{g}\sigma_{x}(a^{\dagger}-a)/\omega_{\mathrm{c}}}$. 
We recall the fact that the quantum Rabi Hamiltonian 
$H\left(\omega_{\mathrm{a}},0,\omega_{\mathrm{c}},\mathrm{g}\right)$ 
has the parity symmetry, i.e., 
$[H\left(\omega_{\mathrm{a}},0,\omega_{\mathrm{c}},\mathrm{g}\right), 
\Pi]=0$ for the parity operator 
$\Pi=(-1)^{a^{\dagger}a}\sigma_{z}$, 
and the ground state $|E_{0}^{\varepsilon=0}\rangle$ of the quantum Rabi Hamiltonian 
has to be an eigenstate of the parity operator, 
of which eigenvalue is $-1$, i.e., 
$\Pi|E_{0}^{\varepsilon=0}\rangle=-|E_{0}^{\varepsilon=0}\rangle$ 
\cite{HHL,hiro-hiro}. 
Thus, this convergence say that for sufficiently large coupling strength, $\mathrm{g}\gg 1$, 
the ground state $|E_{0}^{\varepsilon=0}\rangle$ of the quantum Rabi Hamiltonian 
$H\left(\omega_{\mathrm{a}},0,\omega_{\mathrm{c}},\mathrm{g}\right)$ 
is well approximated by the ground state 
$U_{\mathrm{g}}^{*}|\downarrow, 0\rangle$ 
of the Hamiltonian 
$U_{\mathrm{g}}^{*}H_{\mathrm{cavity}}(\omega_{\mathrm{c}})
U_{\mathrm{g}}$. 
Namely, 
the ground state $U_{xz}|E_{0}^{\varepsilon=0}\rangle$ 
of the original total Hamiltonian $\mathcal{H}_{\mathrm{total}}$ 
is well approximated by the coherent states, 
also called the displaced vacuum states \cite{semba2016}, 
as 
\begin{eqnarray}
U_{xz}|E_{0}^{\varepsilon=0}\rangle
&\approx& 
\frac{
\hat{D}(-\mathrm{g}/\omega_{\mathrm{c}})|\uparrow, 0\rangle
+\hat{D}(\mathrm{g}/\omega_{\mathrm{c}})|\downarrow, 0\rangle
}{\sqrt{2}} 
\nonumber \\ 
&=& 
\frac{e^{-\mathrm{g}^{2}/\omega_{\mathrm{c}}^{2}}}{\sqrt{2}}
\sum_{n=0}^{\infty}
\left\{
\frac{(-\mathrm{g}/\omega_{\mathrm{c}})^{n}}{\sqrt{n!}}
|\uparrow,n\rangle
+
\frac{(\mathrm{g}/\omega_{\mathrm{c}})^{n}}{\sqrt{n!}}
|\downarrow,n\rangle
\right\}
\label{eq:approximation_coherent_state}
\end{eqnarray}
for the displacement operator 
$\hat{D}(\mp\mathrm{g}/\omega_{\mathrm{c}})
=e^{\mp\mathrm{g}(a^{\dagger}-a)/\omega_{\mathrm{c}}}$. 
This improvement of convergence also guarantees the convergence 
of energy spectra as $\mathrm{g}\to\infty$, 
and our numerical analysis in, 
for instance, Fig.\ref{fig:energy_rabi}. 
It is worth to noting that Braak has given an analytical expression of the eigenvalues 
of the quantum Rabi Hamiltonian \cite {braak1} 
though it had been a long outstanding problem.  

Here, we give a proof of the improvement of the convergence. 
Denote $H_{\mathrm{cavity}}\left(\omega_{\mathrm{c}}\right)$ by 
$H_{\mathrm{cav}}$ for simplicity. 
We have 
$U_{\mathrm{g}}aU_{\mathrm{g}}^{*}
=
a-\mathrm{g}\sigma_{x}/\omega_{\mathrm{c}}$. 
Then, we can define a Hamiltonian $\widetilde{H}^{\mathrm{ren}}_{\mathrm{Rabi}}$ by 
\begin{equation}
\widetilde{H}^{\mathrm{ren}}_{\mathrm{Rabi}}:=
U_{\mathrm{g}}\left(
H_{\mathrm{Rabi}}\left(\omega_{\mathrm{a}},0,\omega_{\mathrm{c}},\mathrm{g}\right)
+\frac{\hbar\mathrm{g}^{2}}{\omega_{\mathrm{c}}}
\right)U_{\mathrm{g}}^{*}
=H_{\mathrm{cav}}+W_{\mathrm{g}},  
\label{eq:improve-1}
\end{equation}
where  
$W_{\mathrm{g}}=(\hbar\omega_{\mathrm{a}}/2)U_{\mathrm{g}}\sigma_{z}
U_{\mathrm{g}}^{*}$. 
Set the operator $R$ as $R:=(\widetilde{H}^{\mathrm{ren}}_{\mathrm{Rabi}}-i)^{-1}
-
\left( H_{\mathrm{cav}}-i\right)^{-1}$, and use the $2$nd resolvent equation. 
Then, we have the equation, 
\begin{equation}
R=
\left( H_{\mathrm{cav}}+W_{\mathrm{g}}-i\right)^{-1}
\left( -W_{\mathrm{g}}\right)
\left( H_{\mathrm{cav}}-i\right)^{-1}, 
\label{eq:improve-2}
\end{equation}
which implies the equation, 
\begin{equation}
\left( H_{\mathrm{cav}}+W_{\mathrm{g}}-i\right)^{-1}
=
\left( H_{\mathrm{cav}}-i\right)^{-1} 
+\left( H_{\mathrm{cav}}+W_{\mathrm{g}}-i\right)^{-1}
\left( -W_{\mathrm{g}}\right)
\left( H_{\mathrm{cav}}-i\right)^{-1}.
\label{eq:improve-3}
\end{equation}
Inserting Eq.(\ref{eq:improve-3}) into the term, 
$\left( H_{\mathrm{cavh}}+W_{\mathrm{g}}-i\right)^{-1}$, 
of the right hand side of Eq.(\ref{eq:improve-2}), 
we reach the equation, 
\begin{eqnarray*}
R&=&
-\left( H_{\mathrm{cav}}-i\right)^{-1}
W_{\mathrm{g}}
\left( H_{\mathrm{cav}}-i\right)^{-1} \\ 
&{}&
+
\left( H_{\mathrm{cav}}+W_{\mathrm{g}}-i\right)^{-1}
W_{\mathrm{g}}
\left( H_{\mathrm{cav}}-i\right)^{-1}
W_{\mathrm{g}}
\left( H_{\mathrm{cav}}-i\right)^{-1}, 
\end{eqnarray*}
which implies the inequality, 
\begin{eqnarray*}
\| R\|_{\mathrm{op}}
&\le& 
\| \left( H_{\mathrm{cav}}-i\right)^{-1}
W_{\mathrm{g}}
\left( H_{\mathrm{cav}}-i\right)^{-1}\|_{\mathrm{op}} \\ 
&{}&+
\| \left( H_{\mathrm{cav}}
+W_{\mathrm{g}}-i\right)^{-1}\|_{\mathrm{op}}
\| W_{\mathrm{g}}\|_{\mathrm{op}}
\|\left( H_{\mathrm{cav}}-i\right)^{-1}
W_{\mathrm{g}}
\left( H_{\mathrm{cav}}-i\right)^{-1}\|_{\mathrm{op}}.
\end{eqnarray*}
Since $\| \left( H_{\mathrm{cav}}
+W_{\mathrm{g}}-i\right)^{-1}\|_{\mathrm{op}}
\le 1$ and $\| W_{\mathrm{g}}\|_{\mathrm{op}}=\hbar\omega_{\mathrm{a}}/2$, 
the above inequality leads to the inequality 
$$
\| R\|_{\mathrm{op}}
\le 
\left( 1+\frac{\hbar\omega_{\mathrm{a}}}{2}\right)
\|\left( H_{\mathrm{cav}}-i\right)^{-1}
W_{\mathrm{g}}
\left( H_{\mathrm{cav}}-i\right)^{-1}\|_{\mathrm{op}}.
$$ 
We remember that the operator $W_{\mathrm{g}}$ converges to $0$ 
in the weak operator topology \cite{hiro-qs}, 
and the resolvent $\left( H_{\mathrm{cav}}-i\right)^{-1}$ is compact. 
Our theorem proved in \ref{app:theorem} says that 
$\lim_{\mathrm{g}\to\infty}\|\left( 1+\left(\hbar\omega_{\mathrm{a}}/2\right)\right)
\|\left( H_{\mathrm{cav}}-i\right)^{-1}
W_{\mathrm{g}}
\left( H_{\mathrm{cav}}-i\right)^{-1}\|_{\mathrm{op}}=0$, 
and therefore, 
\begin{eqnarray*}
&{}&\lim_{\mathrm{g}\to\infty}
\| 
\left(H_{\mathrm{Rabi}}\left(\omega_{\mathrm{a}},\omega_{\mathrm{c}},\mathrm{g}\right)
+\hbar\mathrm{g}^{2}/\omega_{\mathrm{c}}
-i\right)^{-1}
-
\left( U_{\mathrm{g}}^{*}H_{\mathrm{cav}}U_{\mathrm{g}}-i\right)^{-1}
\|_{\mathrm{op}}  \\ 
&=& \lim_{\mathrm{g}\to\infty}
\Bigl\| \left( U_{\mathrm{g}}
H_{\mathrm{Rabi}}\left(\omega_{\mathrm{a}},\omega_{\mathrm{c}},\mathrm{g}\right)
U_{\mathrm{g}}^{*}+\hbar\mathrm{g}^{2}/\omega_{\mathrm{c}}
-i\right)^{-1}
-
\left( H_{\mathrm{cav}}-i\right)^{-1}
\Bigr\|_{\mathrm{op}} 
=0. 
\end{eqnarray*}

Based on this convergence and its proof, 
we think that the leading term of the quantum Rabi Hamiltonian 
is $H_{\mathrm{cavity}}(\omega_{\mathrm{c}})+\hbar\mathrm{g}(\sigma_{+}a+\sigma_{-}a^{\dagger})$ 
for ultra- or deep-strong coupling regime. 
Namely, we regard $H_{\mathrm{atom}}(\omega_{\mathrm{a}},0)$ 
as a small perturbation. 

\section{Dressed Photon in Ground State}

\begin{center}
\begin{figure*}
\begin{center}
\includegraphics[width=0.3\textwidth]{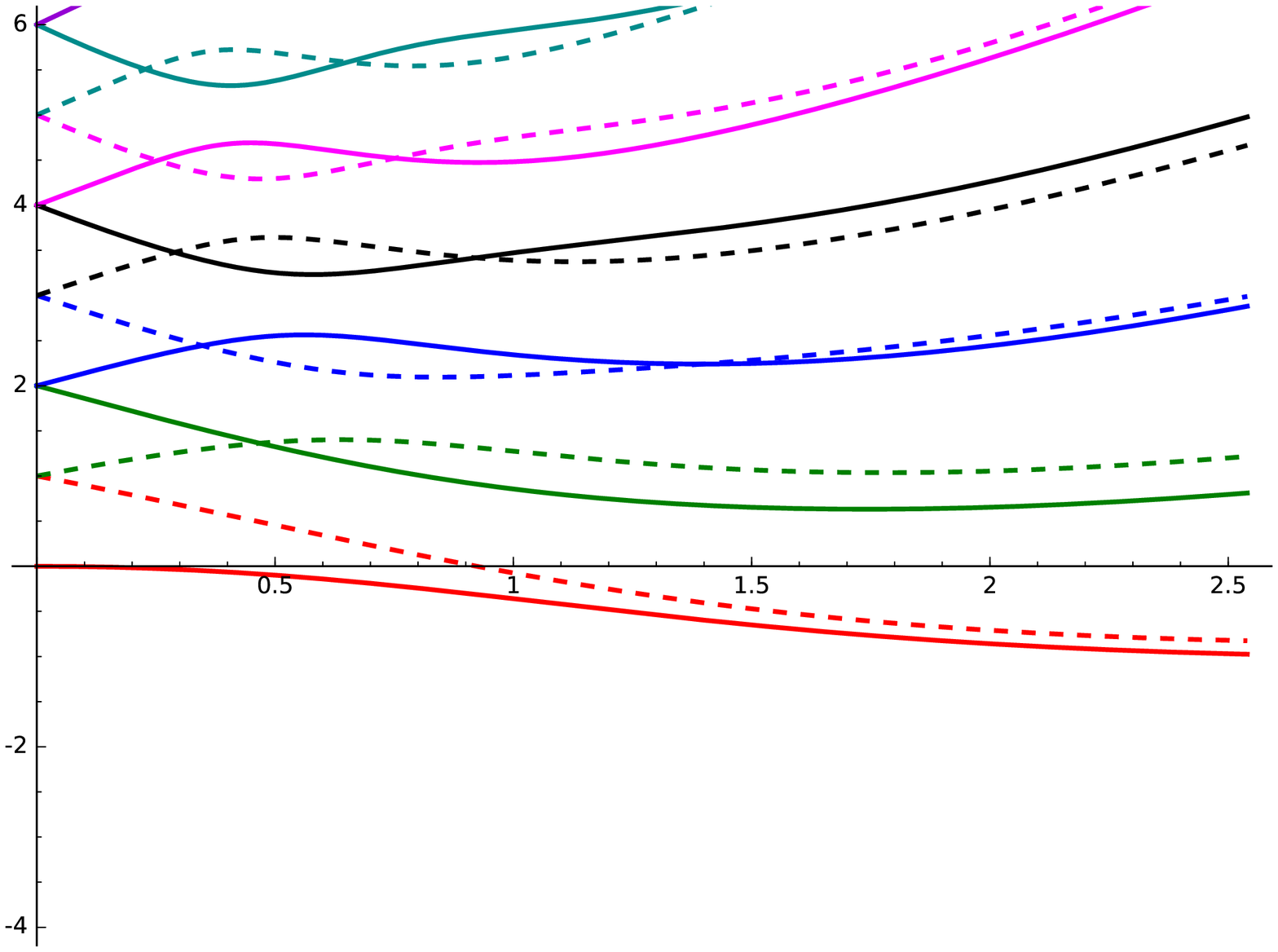}
\qquad\qquad 
\includegraphics[width=0.35\textwidth]{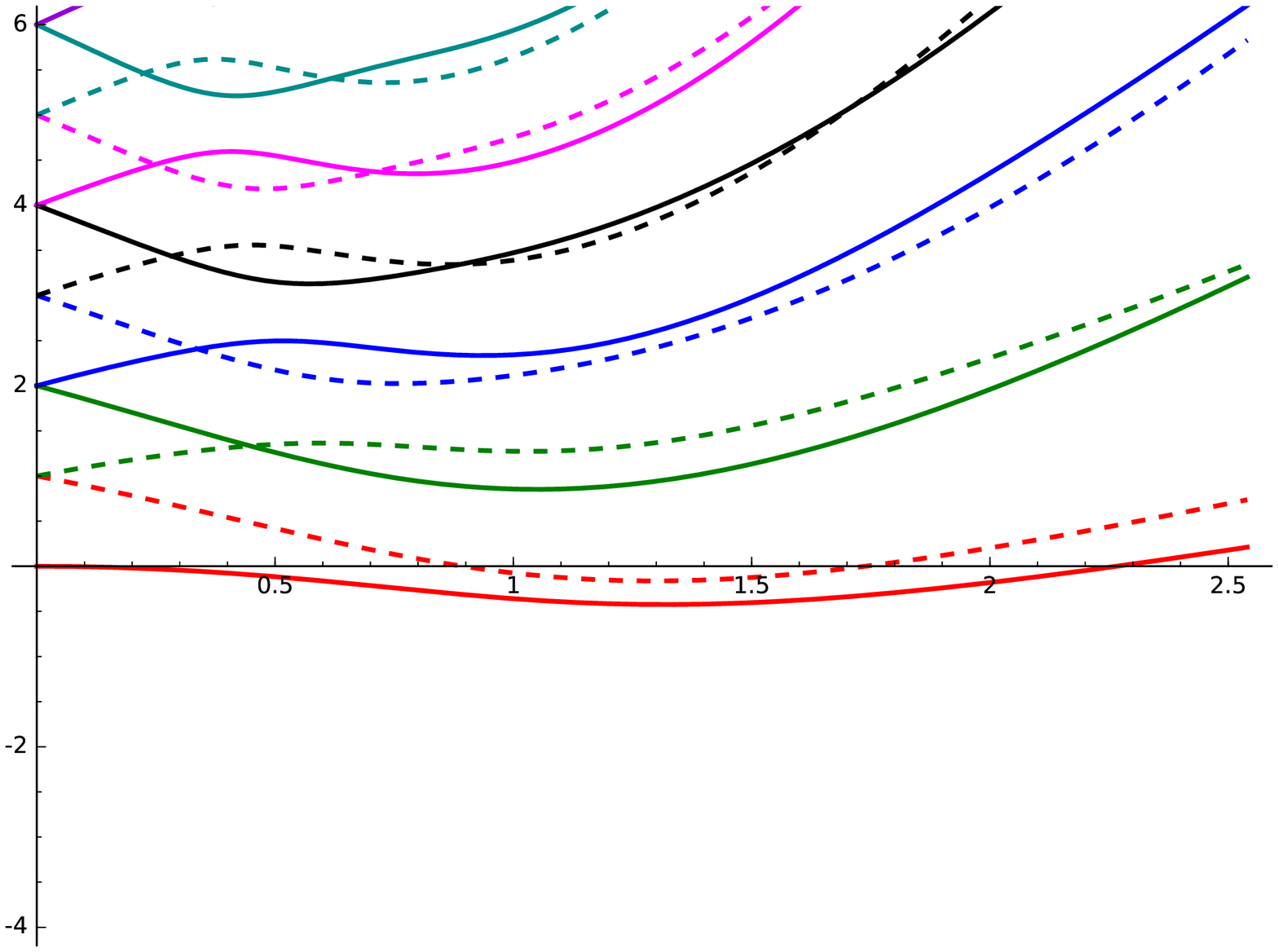}
\vspace*{12mm}
\caption{
The energy revel of the total Rabi Hamiltonian 
$H_{A^{2}}$. 
The left is for the case $C_{\mathrm{g}}=0.1\times\mathrm{g}$, 
and the right for the case $C_{\mathrm{g}}=0.1\times\mathrm{g}^{2}$. 
The unit of the transverse axis is $\mathrm{g}/\omega$, 
and the unit of the longitudinal axis is $\hbar\omega$.
} 
\label{fig:energy_rabi_A^2}       % Give a unique label
\end{center}
\end{figure*}
\end{center}

In this section we will consider the dressed photon in 
a ground state of the total Hamiltonian with 
the $A^{2}$-term, the quadratic coupling, 
defined by  
\begin{equation}
H_{A^{2}}
:=H(\omega_{\mathrm{a}},\varepsilon,\omega_{\mathrm{c}},\mathrm{g})+\hbar C_{\mathrm{g}}\mathrm{g}\left(a+a^{\dagger}\right)^{2}, 
\end{equation}
where $C_{\mathrm{g}}$ is a function of the coupling strength $\mathrm{g}$. 
Judging from reference to Eq.(15) of Ref.\cite{semba2016}, the coupling strength $\mathrm{g}$ 
and the function $C_{\mathrm{g}}$ are given by 
$$
\mathrm{g}=M(I_{\mathrm{p}})I_{\mathrm{p}}I_{\mathrm{zpf}}\quad
\textrm{and}\quad 
C_{\mathrm{g}}=
\frac{I_{\mathrm{p}}I_{\mathrm{zpf}}}{I_{\mathrm{cM}}^{2}-I_{\mathrm{p}}^{2}}, 
$$
where $I_{\mathrm{p}}$ and $I_{\mathrm{zpf}}$ are respectively 
the persistent-current and the zero-point-fluctuation current 
with $I_{\mathrm{zpf}} \ll I_{\mathrm{p}}$,  
$I_{\mathrm{cM}}$ is the critical current of a single effective Josephson junction, 
and $M(I_{\mathrm{p}})$ the mutual inductance between the flux qubit and the LC oscillator:
$$
M(I_{\mathrm{p}})=\frac{\Phi_{0}}{2\pi\sqrt{I_{\mathrm{cM}}^{2}-I_{\mathrm{p}}^{2}}}.
$$
We can rewrite $C_{\mathrm{g}}$ as 
$$
C_{\mathrm{g}}=
\frac{I_{\mathrm{p}}I_{\mathrm{zpf}}}{I_{\mathrm{cM}}^{2}-I_{\mathrm{p}}^{2}} 
=\frac{2\pi}{\Phi_{0}\sqrt{I_{\mathrm{cM}}^{2}-I_{\mathrm{p}}^{2}}}\mathrm{g}
\qquad 
\textrm{or}\qquad 
C_{\mathrm{g}}=
\frac{I_{\mathrm{p}}I_{\mathrm{zpf}}}{I_{\mathrm{cM}}^{2}-I_{\mathrm{p}}^{2}} 
=\left(\frac{2\pi}{\Phi_{0}}\right)^{2}
\frac{1}{I_{\mathrm{p}}I_{\mathrm{zpf}}}\mathrm{g}^{2}.
$$
Thus, we may approximate it at $C_{\mathrm{g}}=C\mathrm{g}^{\ell}$, 
$\ell=1, 2$, with a constant $C$, 
where we have to change the dimension of $C$ to meet each case. 
In the latter case, $C\ll 1$ as $\mathrm{g}\gg 1$ 
if we take sufficiently large $I_{\mathrm{p}}I_{\mathrm{zpf}}$ 
to make $\mathrm{g}$ large enough. 
The numerical computation of the energy level of the total Hamiltonian $H_{A^{2}}$ 
is in Fig.\ref{fig:energy_rabi_A^2}. 

We denote by the eigenstates $|E_{\nu}\rangle$, $\nu=0, 1, 2, \cdots$, 
of the total Hamiltonian 
$H_{A^{2}}$, and by $E_{\nu}$ the corresponding eigenenergies 
with $E_{0}\le E_{1}\le E_{2}\le \cdots$. 
Return the representation used by Yoshihara \textit{et al}. \cite{semba2016}, 
\begin{equation}
U_{xz}H_{A^{2}}U_{xz}^{*}
=\mathcal{H}_{\mathrm{total}}+\hbar C_{\mathrm{g}}\mathrm{g}
(a+a^{\dagger})^{2}.
\label{eq:original-representation}
\end{equation}
Then, we have the matrix representation, 
$$
U_{xz}H_{A^{2}}U_{xz}^{*}
=
\left(
\begin{array}{cccc}
H_{A^{2}}^{+}-\hbar\varepsilon/2 & -\hbar\omega_{\mathrm{a}}/2 \\ 
-\hbar\omega_{\mathrm{a}}/2 & H_{A^{2}}^{-}+\hbar\varepsilon/2
\end{array}
\right), 
$$
where 
$$
H_{A^{2}}^{\pm}=
H_{\mathrm{cavity}}(\omega_{\mathrm{c}})
\pm\hbar\mathrm{g}\left( a+ a^{\dagger}\right) 
+\hbar C_{\mathrm{g}}\mathrm{g}
\left( a+ a^{\dagger}\right)^{2}. 
$$
In other words, following the representation 
in Ref.\cite{casanoba-solano}, we define 
the annihilation operator $\alpha$ 
and creation operator $\alpha^{\dagger}$ by 
$\alpha^{\sharp}:=\sigma_{x}a^{\sharp}$. 
Then, we have the matrix-valued CCR, $[\alpha , \alpha^{\dagger}]=1$. 
We denote by $\Pi_{\pm 1}$ the orthogonal projection on the eigenspace of the 
parity operator $\Pi$ with the corresponding eigenvalue $\pm 1$. 
Then, we have the relations, $\Pi=\, -\Pi_{-1}+\Pi_{+1}$ 
and $1=\Pi_{-1}+\Pi_{+1}$. 
Using these relations, we have the well-known representation, 
\begin{equation}
H_{A^{2}}=H_{-}\Pi_{-1}\oplus H_{+}\Pi_{+1} 
-\frac{\hbar\varepsilon}{2}\sigma_{x},
\label{eq:parity-decomposition}
\end{equation}
where $H_{-}$ and $H_{+}$ are the photon Hamiltonians given by 
\begin{equation}
H_{\pm}=\hbar\omega_{\mathrm{c}}\left(\alpha^{\dagger}\alpha
+\frac{1}{2}\right)
+\hbar\mathrm{g}(\alpha+\alpha^{\dagger})
+\hbar C_{\mathrm{g}}\mathrm{g}
(\alpha+\alpha^{\dagger})^{2} 
\pm\frac{\hbar\omega_{\mathrm{a}}}{2}(-1)^{\alpha^{\dagger}\alpha}. 
\label{eq:parity-decomposition2}
\end{equation}
We regard the term,  
$\hbar\omega_{\mathrm{c}}\left(\alpha^{\dagger}\alpha
+1/2\right)
+\hbar\mathrm{g}(\alpha+\alpha^{\dagger})
+\hbar C_{\mathrm{g}}\mathrm{g}
(\alpha+\alpha^{\dagger})^{2}$, 
with the self-interaction 
as the leading term, and the terms, 
$\pm(\hbar\omega_{\mathrm{a}}/2)(-1)^{\alpha^{\dagger}\alpha}$ 
and $-(\hbar\varepsilon/2)\sigma_{x}$, 
as its perturbation as $\mathrm{g}\gg 1$. 

According to the argument on p.76 and p.77 of Ref.\cite{henley-thirring}, 
the ground-state expectation of photon defined by 
$$
N_{0}^{\mathrm{bare}}
=\langle E_{0}|a^{\dagger}a|E_{0}\rangle
=\langle E_{0}|\alpha^{\dagger}\alpha|E_{0}\rangle
$$ 
is for the bare photons.  
Thus, from the point of view of the photon field with the uncountably many modes, 
the ground-state expectation $N_{0}^{\mathrm{bare}}$ includes the number 
of virtual photons as well as that of non-virtual photons. 
Here `virtual photon' is a technical term 
in the theory of elementary particle, and `non-virtual' is its 
antonym of the virtual though some people use `real' instead of non-virtual. 
The virtual photon causes the Coulomb force in quantum electrodynamics, 
though it cannot directly be observed in any experiment. 
Thus, we should investigate the effect coming from the fluctuation 
caused by bare photons. 

Let $\Delta\Phi$ be the fluctuation of the bare-photon field 
$\Phi=(a+a^{\dagger})/\sqrt{2\omega_{\mathrm{c}}}$ 
at the ground state $|E_{0}\rangle$, i.e., 
$\Delta\Phi=\sqrt{\langle E_{0}|(\Phi-\langle E_{0}|\Phi|E_{0}\rangle)^{2}|E_{0}\rangle}$. 
We can show the relation, 
\begin{equation}
(\Delta\Phi)^{2}\le 
\frac{2N_{\mathrm{0}}^{\mathrm{bare}}+1}{\omega_{\mathrm{c}}}. 
\label{eq:field_fluctuation-bare}
\end{equation}
In addition, according to the way in Eq.(12.24) of Ref.\cite{henley-thirring}, 
we can obtain the following lower bound 
in the case $C_{\mathrm{g}}=C\mathrm{g}^{\ell}$, 
$\ell=1, 2$, with a constant $C>0$: 
\begin{equation}
N_{0}^{\mathrm{bare}}
\ge 
\frac{1}{2}
\left(
\frac{\sqrt{\omega_{\mathrm{c}}}}{2\sqrt{
\omega_{\mathrm{c}}+
4C\mathrm{g}^{\ell+1}}}
+
\frac{\sqrt{
\omega_{\mathrm{c}}+
4C\mathrm{g}^{\ell+1}}}{2\sqrt{\omega_{\mathrm{c}}}}
-1
\right)
-\epsilon(\mathrm{g}),
\label{eq:bare_lower_bnd} 
\end{equation}
where 
$$
0<\epsilon(\mathrm{g})\to 
\left\{
\begin{array}{ll}
1/(8C\omega_{\mathrm{c}}) & \textrm{$\ell=1$}, \\ 
0 & \textrm{$\ell=2$}, 
\end{array}
\right.
\textrm{as $\mathrm{g}\to\infty$}.
$$ 
We will give the proofs of Eqs.(\ref{eq:field_fluctuation-bare}) and 
(\ref{eq:bare_lower_bnd}) in \ref{app:bare}. 
%Particularly, in the case $C_{\mathrm{g}}=0$, making the coupling strength larger 
%(i.e., $\mathrm{g}\to\infty$), 
Thus, we have to avoid the increase in the number of bare photons 
which tends to divergence. 
Also, we have to mind and cope with the effect coming from the quadratic coupling. 
Therefore, we will follow the pair theory \cite{henley-thirring} to consider the physical state. 
The pair theory is primarily to expose non-virtual photons to us 
by removing virtual photons from the ground state.

\subsection{Pair Theory for $H_{A^{2}}$} 
\label{sec:eA2}

In this part, we mathematically study the Hopfield-Bogoliubov transformation 
to cope with the quadratic coupling. 

Following Eqs.(12.17) and (12.19) of Ref.\cite{henley-thirring}, 
the annihilation and creation operators, $b$ and $b^{\dagger}$, 
for physical photons 
are introduced so that the relations, 
\begin{equation}
a=M_{1}b+M_{2}b^{\dagger}
\quad\textrm{and}\quad 
a^{\dagger}=M_{2}b+M_{1}b^{\dagger},
\label{eq:result0}
\end{equation}
are satisfied, where 
$$
M_{1}=\frac{1}{2}\left\{
\sqrt{\frac{\omega_{\mathrm{c}}}{\omega_{\mathrm{g}}}}
+\sqrt{\frac{\omega_{\mathrm{g}}}{\omega_{\mathrm{c}}}}
\right\}\quad\textrm{and}\quad 
M_{2}=\frac{1}{2}\left\{
\sqrt{\frac{\omega_{\mathrm{c}}}{\omega_{\mathrm{g}}}}
-\sqrt{\frac{\omega_{\mathrm{g}}}{\omega_{\mathrm{c}}}}
\right\}.
$$  
Then, we can prove that for arbitrary 
$\omega_{\mathrm{c}}, \mathrm{g}, C_{\mathrm{g}}$ 
there is a unitary operator $U$ such that 
\begin{equation}
U^{*}H_{A^{2}}U
=H(\omega_{\mathrm{a}},\varepsilon,\omega_{\mathrm{g}},\widetilde{\mathrm{g}}), 
\label{eq:result}
\end{equation}
where 
$$
\omega_{\mathrm{g}}
=\sqrt{\omega_{\mathrm{c}}^{2}+4C_{\mathrm{g}}\mathrm{g}\omega_{\mathrm{c}}}\quad 
\textrm{and}\quad 
\widetilde{\mathrm{g}}=\mathrm{g}\sqrt{\frac{\omega_{\mathrm{c}}}{\omega_{\mathrm{g}}}}.
$$
This unitary operator $U$ is the Hopfield-Bogoliubov 
transformation \cite{hopfield,ciuti} used in Ref. \cite{semba2016}. 
We here note that this Hopfield-Bogoliubov transformation becomes unitary 
without any restriction. 
The cavity frequency  and the coupling strength are respectively renormalized 
as $\omega_{\mathrm{g}}$ and $\widetilde{\mathrm{g}}$ 
so that they include the effect 
coming from the $A^{2}$-term. 
In particular, the term, $2\sqrt{C_{\mathrm{g}}\mathrm{g}\omega_{\mathrm{c}}}$, 
in the renormalized cavity frequency $\omega_{\mathrm{g}}$ plays a role of 
a mass of photon (cf. \S 3.4 of Ref.\cite{preparata}).   
Thus, we call the Hamiltonian 
$H(\omega_{\mathrm{a}},\varepsilon,\omega_{\mathrm{g}}, \widetilde{\mathrm{g}})$ 
the renormalized Hamiltonian with $A^{2}$-term effect. 

Using this unitarily-equivalent representation, 
in the same way as getting Eq.(1.18) of Ref.\cite{hiro-rmp}, 
we can estimate the ground state energy 
$E_{0}(H_{A^{2}})$ of the Hamiltonian $H_{A^{2}}$ as 
\begin{equation}
-\, \frac{\hbar}{2}\sqrt{\omega_{\mathrm{a}}^{2}+\varepsilon^{2}}
+\frac{\hbar\omega_{\mathrm{g}}}{2}
-\, \frac{{\hbar\widetilde{\mathrm{g}}}^{2}}{\omega_{\mathrm{g}}}
\le 
E_{0}(H_{A^{2}})
\le 
-\, \frac{\hbar}{2}
e^{-2{\widetilde{\mathrm{g}}}^{2}/{\omega_{\mathrm{g}}}^{2}}
\sqrt{\omega_{\mathrm{a}}^{2}+\varepsilon^{2}}
+\frac{\hbar\omega_{\mathrm{g}}}{2}
-\, \frac{\hbar{\widetilde{\mathrm{g}}}^{2}}{\omega_{\mathrm{g}}}. 
\label{eq:gse-1}
\end{equation}

We can show how the Hopfield-Bogoliubov transformation is obtained as a unitary operator. 
We are now considering just one-mode photon, 
not uncountably many modes light. 
Thus, we only have to consider the Schr\"{o}dinger equation,  
in particular, the Schr\"{o}dinger operators in our case 
instead of considering the field equation in pair theory: 
Return the representation $U_{xz}H_{A^{2}}U_{xz}^{*}$ 
in Eq.(\ref{eq:original-representation}) again. 
We recall that the unitary operator $U_{xz}$ is only for the spin, 
and our Hopfield-Bogoliubov transformation $U$ only for photon. 
We can represent the annihilation operator $a$ and the creation operator $a^{\dagger}$ 
of bare photons using the position operator $x$ and the momentum operator $p$ as 
\begin{equation}
a=\sqrt{\frac{m\omega_{\mathrm{c}}}{2\hbar}}\, x
+i\sqrt{\frac{1}{2m\hbar\omega_{\mathrm{c}}}}\, p  
\quad\textrm{and}\quad
a^{\dagger}=\sqrt{\frac{m\omega_{\mathrm{c}}}{2\hbar}}\, x
-i\sqrt{\frac{1}{2m\hbar\omega_{\mathrm{c}}}}\, p. 
\label{eq:pm_a}
\end{equation}
Use these representations, and rewrite the Hamiltonians $H_{A^{2}}^{\pm}$ as 
\begin{eqnarray*}
H_{A^{2}}^{\pm}&=&
\frac{1}{2m}p^{2}+\frac{m\omega_{\mathrm{c}}^{2}}{2}x^{2}
\pm\hbar\mathrm{g}\sqrt{\frac{2m\omega_{\mathrm{c}}}{\hbar}}\, x
+2mC_{\mathrm{g}}\mathrm{g}\omega_{\mathrm{c}}x^{2} \\ 
&=& 
\frac{1}{2m}p^{2}+\frac{m\omega_{\mathrm{g}}^{2}}{2}x^{2}
\pm\hbar\widetilde{\mathrm{g}}\sqrt{\frac{2m\omega_{\mathrm{g}}}{\hbar}}\, x. 
\end{eqnarray*} 
We denote by $\varphi_{n}$ the normalized eigenstates of 
$(1/2m)p^{2}+(m\omega_{\mathrm{c}}^{2}/2)x^{2}$, 
and by $\psi_{n}$ the normalized eigenstates of 
$(1/2m)p^{2}+(m\omega_{\mathrm{g}}^{2}/2)x^{2}$ 
for $n=0, 1, 2, \cdots$:
$$
\varphi_{n}=\sqrt{w_{\mathrm{c}}}\gamma_{n}
H_{n}(w_{\mathrm{c}}x)e^{-(w_{\mathrm{c}}x)^{2}/2}\quad 
\textrm{and}\quad
\psi_{n}=\sqrt{w_{\mathrm{g}}}\gamma_{n}
H_{n}(w_{\mathrm{g}}x)e^{-(w_{\mathrm{g}}x)^{2}/2},
$$
where $H_{n}(x)$ are the Hermite polynomials of variable $x$, 
$\gamma_{n}=\pi^{-1/4}(2^{n}n!)^{-1/2}$, 
and $w_{\sharp}=\sqrt{m\omega_{\sharp}/\hbar}$, 
$\sharp=\mathrm{c}, \mathrm{g}$. 
Since the sets, $\left\{\varphi_{n}\right\}_{n=0}^{\infty}$ 
and $\left\{\psi_{n}\right\}_{n=0}^{\infty}$, 
are respectively a complete orthonormal basis of the function space 
$L^{2}(\mathbb{R})$,  we can make a unitary operator $U$ 
by the correspondence, $U\varphi_{n}=\psi_{n}$. 
We here note the relations, $a\varphi_{0}=0$ and 
$a\varphi_{n}=\sqrt{n}\varphi_{n-1}$ for $n=1, 2, \cdots$.

Introducing the annihilation and creation operators, $b$ and $b^{\dagger}$, 
for physical photons by 
\begin{equation}
b=\sqrt{\frac{m\omega_{\mathrm{g}}}{2\hbar}}\, x
+i\sqrt{\frac{1}{2m\hbar\omega_{\mathrm{g}}}}\, p 
\quad\textrm{and}\quad  
b^{\dagger}=\sqrt{\frac{m\omega_{\mathrm{g}}}{2\hbar}}\, x
-i\sqrt{\frac{1}{2m\hbar\omega_{\mathrm{g}}}}\, p, 
\label{eq:pm_b}
\end{equation}
we have the relations, $b\psi_{0}=0$ and $b\psi_{n}=\sqrt{n}\psi_{n-1}$ 
for $n= 1, 2, \cdots$. 
Meanwhile, the equations, $Ua\varphi_{0}=0$ and $Ua\varphi_{n}=\sqrt{n}U\varphi_{n-1}$, 
lead the equations, $(UaU^{*})\psi_{0}=0$ and $(UaU^{*})\psi_{n}=\sqrt{n}\psi_{n-1}$,  
which implies the relation $UaU^{*}=b$. 
Taking its conjugate, we have the relation, $Ua^{\dagger}U^{*}=b^{\dagger}$.  
Then, using Eqs.(\ref{eq:pm_a}) and (\ref{eq:pm_b}), 
we easily obtain the relations, 
\begin{equation}
\left\{
\begin{array}{ll}
{\displaystyle 
UaU^{*}=b
=\frac{1}{2}(c_{1}+c_{2})a+\frac{1}{2}(c_{1}-c_{2})a^{\dagger}
}, \\ 
\qquad \\ 
{\displaystyle 
Ua^{\dagger}U^{*}=b^{\dagger}
=\frac{1}{2}(c_{1}-c_{2})a+\frac{1}{2}(c_{1}+c_{2})a^{\dagger}
}, 
\end{array}
\right.
\label{eq:4-25-1}
\end{equation}
where 
$c_{1}=\sqrt{\omega_{\mathrm{g}}/\omega_{\mathrm{c}}}$ 
and $c_{2}=\sqrt{\omega_{\mathrm{c}}/\omega_{\mathrm{g}}}$. 
It is easy to show Eqs. (\ref{eq:result0}) and (\ref{eq:result}) with the help of Eqs.(\ref{eq:4-25-1}).

\subsection{Ground-State Expectation of Photons} 
\label{sec:GSEP}

We investigate how the ground state has 
some physical photons that should primarily be 
emitted to the outside of the matter.

Thanks to the unitarity of the Hopfield-Bogoliubov transformation Eq.(\ref{eq:result}), 
we can define the normalized eigenstates of the renormalized total Hamiltonian 
$H(\omega_{\mathrm{a}},\varepsilon,\omega_{\mathrm{g}},\widetilde{\mathrm{g}})$ 
with the $A^{2}$-term effect by $|E_{\nu}^{\mathrm{ren}}\rangle:=
U^{*}|E_{\nu}\rangle$, and then, 
the eigenenergy $E_{\nu}^{\mathrm{ren}}$ of each eigenstate 
$|E_{\nu}^{\mathrm{ren}}\rangle$ is, of course, $E_{\nu}$.  
The result in Ref.\cite{hiro-hiro} says that 
$E_{0}$ is always less than $E_{1}$, 
i.e., 
$E_{0}<E_{1}.$
We consider the renormalized 
ground-state expectation of physical photon, 
$$
N_{0}^{\mathrm{ren}}
=\langle E_{0}^{\mathrm{ren}}|a^{\dagger}a|E_{0}^{\mathrm{ren}}\rangle
=\langle E_{0}|b^{\dagger}b|E_{0}\rangle, 
$$ 
and how the ground state of the renormalized total Hamiltonian 
with the $A^{2}$-term effect has physical photons. 
 
We first find its upper bound as
\begin{equation}
N_{0}^{\mathrm{ren}}\le \frac{\widetilde{\mathrm{g}}^{2}}{\omega_{\mathrm{g}}^{2}}
=\frac{\omega_{\mathrm{c}}}{\omega_{\mathrm{g}}^{3}}\mathrm{g}^{2}
=\omega_{\mathrm{c}}^{-1/2}
\left(
\frac{\omega_{\mathrm{c}}}{\mathrm{g}^{4/3}}
+\frac{4C_{\mathrm{g}}}{\mathrm{g}^{1/3}}
\right)^{-3/2}. 
\label{eq:upp-bnd}
\end{equation}
Eq.(\ref{eq:upp-bnd}), which is proved bellow, 
tells us that how many physical photons 
the ground state can have at most, 
and that it depends on the function $C_{\mathrm{g}}$. 
Thus, we have $N_{0}^{\mathrm{ren}}\to 0$ as $\mathrm{g}\to\infty$ 
in the case $C_{\mathrm{g}}=C\mathrm{g}^{\ell}$, $\ell=1, 2$, while 
$N_{0}^{\mathrm{bare}}\to\infty$. 

Meanwhile, we can give a lower bound of the renormalized 
ground-state expectation. 
Let us define a function $L^{\mathrm{ren}}(\mathrm{g})$ 
of the coupling strength $\mathrm{g}$ 
by 
$$
L^{\mathrm{ren}}(\mathrm{g}):=
\frac{\widetilde{\mathrm{g}}}{\omega_{\mathrm{g}}}\, 
-\sqrt{
\frac{\sqrt{\omega_{\mathrm{a}}^{2}+\varepsilon^{2}}
\left( 
1-e^{-2\widetilde{\mathrm{g}}^{2}/\omega_{\mathrm{g}}^{2}}
\right)}{2\omega_{\mathrm{g}}}}\, .
$$
Then, as proved at the tail end of this subsection, 
for the parameters $\omega_{\mathrm{a}}$, $\varepsilon$, 
$\omega_{\mathrm{c}}$, and $\mathrm{g}$ satisfying 
$L^{\mathrm{ren}}(\mathrm{g})\ge 0$, 
a lower bound is given by 
\begin{equation}
N_{0}^{\mathrm{ren}}\ge L^{\mathrm{ren}}(\mathrm{g})^{2}.
\label{eq:lower-bnd-2}
\end{equation}

\begin{figure*}
\begin{center}
\includegraphics[width=0.4\textwidth]{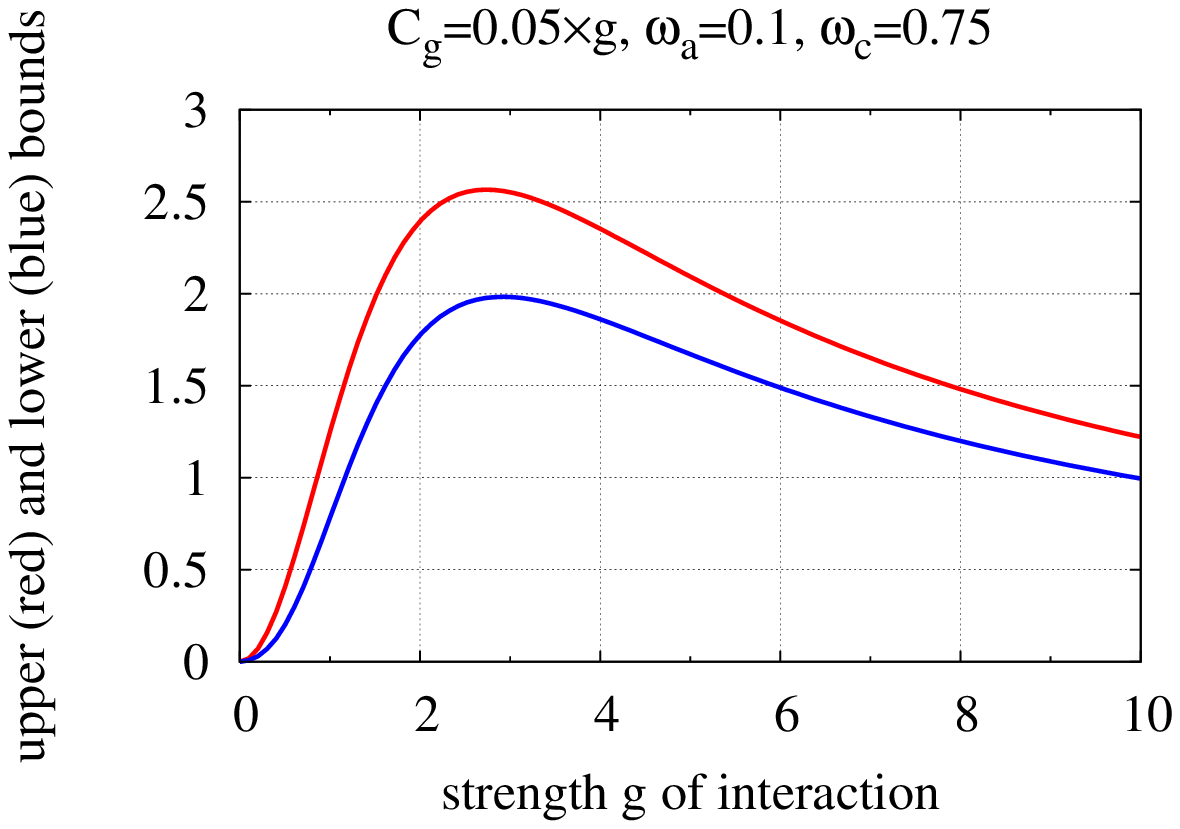}
  \includegraphics[width=0.4\textwidth]{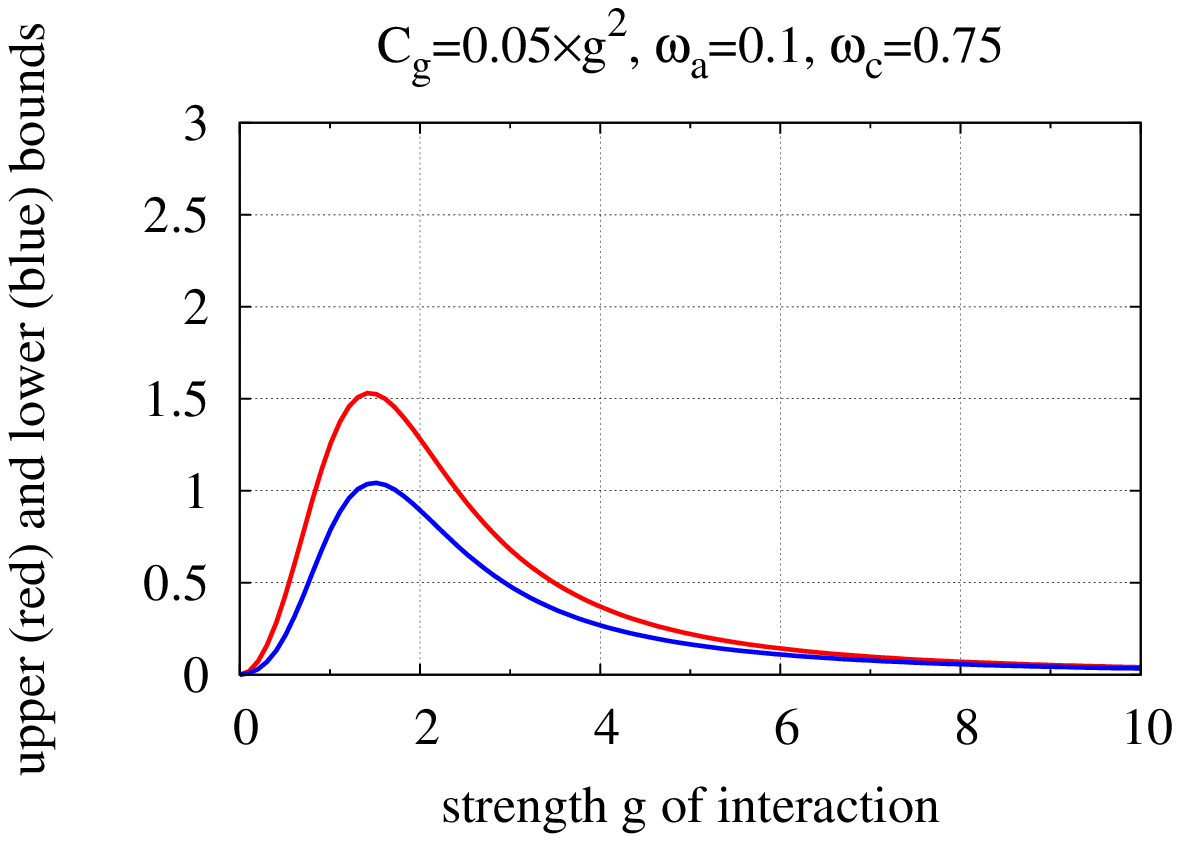} \\   
\includegraphics[width=0.4\textwidth]{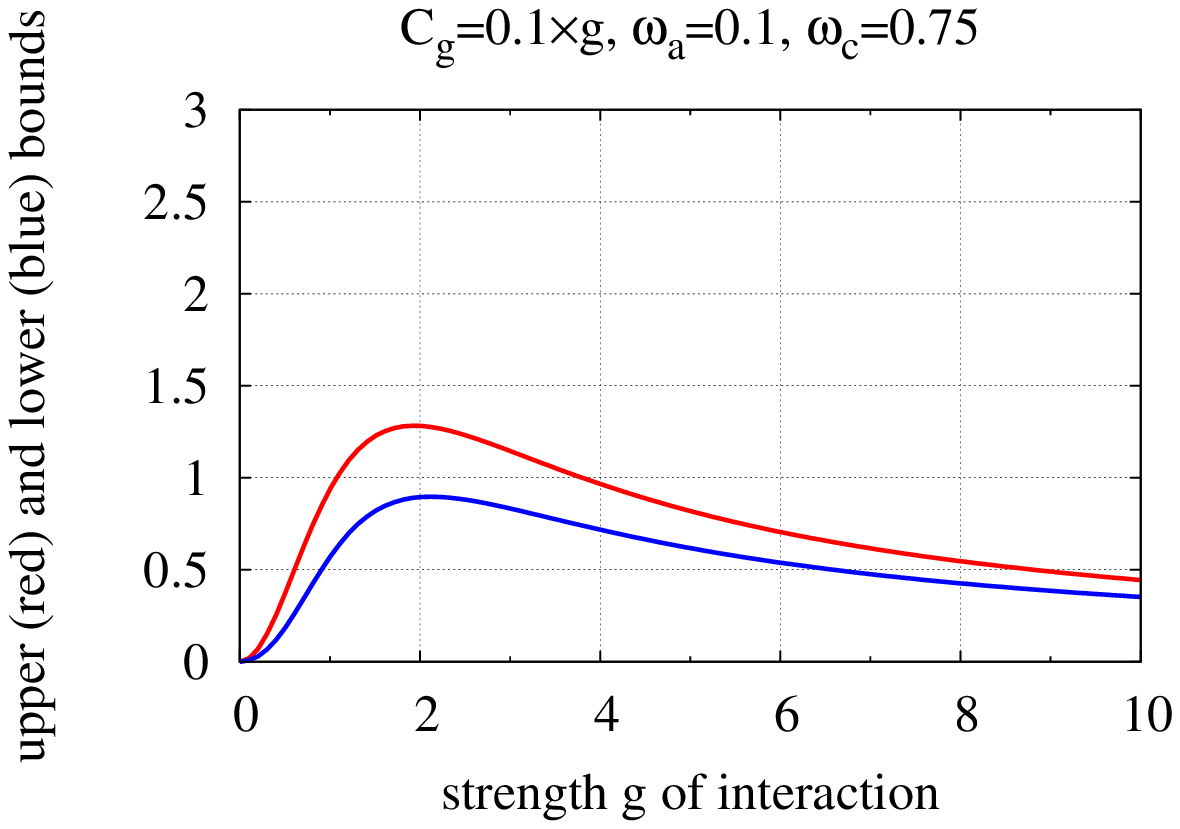}
  \includegraphics[width=0.4\textwidth]{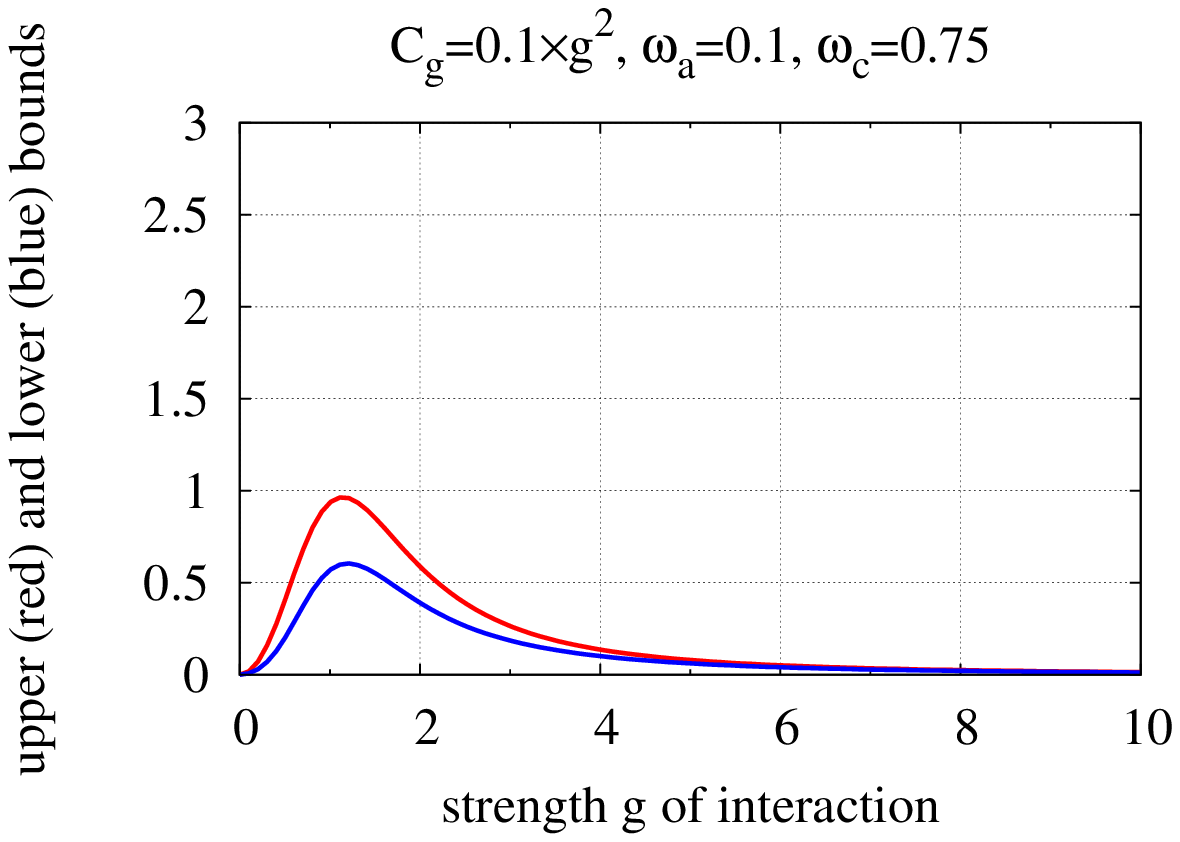} \\ 
\includegraphics[width=0.4\textwidth]{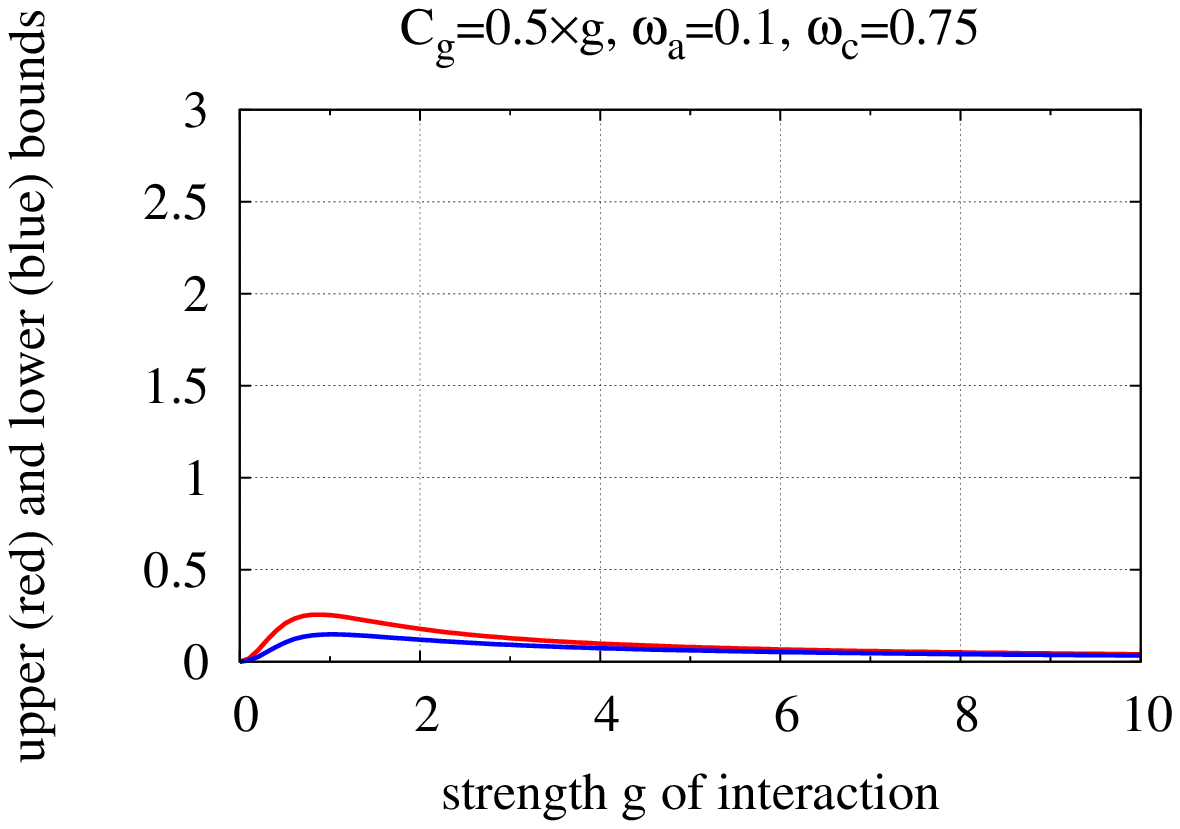}
  \includegraphics[width=0.4\textwidth]{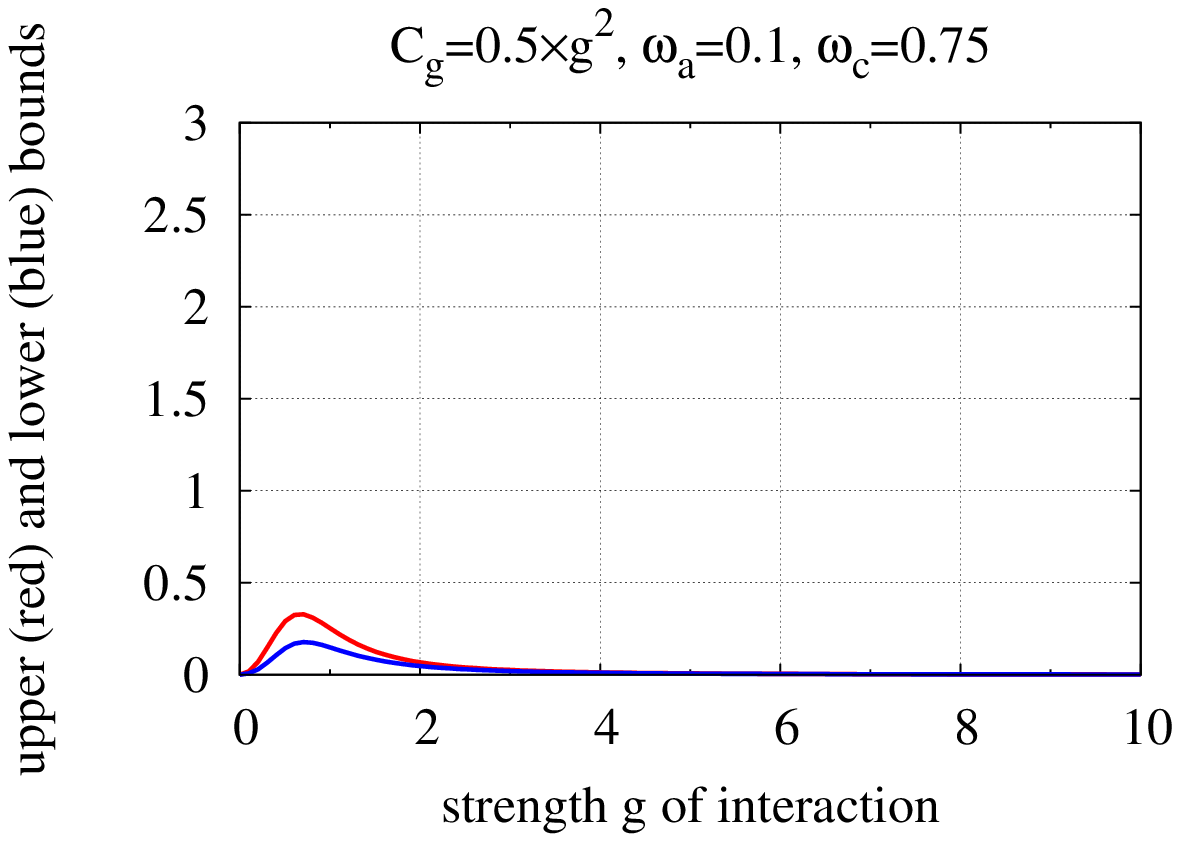}
\vspace*{5mm}
\caption{
The red solid line is the upper bound $\widetilde{\mathrm{g}}^{2}/\omega_{\mathrm{g}}^{2}$, 
and the blue line is the lower bound $L(\mathrm{g})^{2}$. 
Here we set $C_{\mathrm{g}}=C\mathrm{g}$ or $C\mathrm{g}^{2}$, 
$\omega_{\mathrm{a}}=0.1$, $\varepsilon=0$, $\omega_{\mathrm{c}}=0.75$.
} 
\label{fig:upper_lower_bnd_b}       % Give a unique label
\end{center}
\end{figure*}

Now we have the upper bound $\widetilde{\mathrm{g}}^{2}/\omega_{\mathrm{g}}$ 
and the lower bound $L(\mathrm{g})^{2}$. 
Fig.\ref{fig:upper_lower_bnd_b} shows their numerical analyses 
for $C_{\mathrm{g}}=C\mathrm{g}$ or $C\mathrm{g}^{2}$, 
$\omega_{\mathrm{a}}=0.1$, 
$\omega_{\mathrm{c}}=0.75$. 
According to these numerical analyses, 
we realize that the ground-state expectation $N_{0}^{\mathrm{ren}}$ 
has a gently-sloping peak when the constant $C$ is sufficiently small, 
though the peak disappears as the constant is large. 
That is, the hight of the peak is tall for a small $C$, though 
it is low for a large $C$. 
The tall peak tells us that there is a chance so that 
the ground state can have a physical photon.  
This situation reminds us of that of the Hepp-Lieb 
quantum phase transition \cite{ciuti} 
though we have not yet understood whether this is caused by a quantum 
phase transition.

Next our problem is to investigate where the photons that the ground state 
has come from, and how the ground state has them. 
We will hereafter prove the following expression of the renormalized ground-state expectation 
of physical photon. 
For an arbitrary coupling constant $\mathrm{g}$, we have 
\begin{equation}
N_{0}^{\mathrm{ren}}=
|\langle\widetilde{E}_{0}^{\mathrm{ren}}|E_{0}^{\mathrm{ren}}\rangle|^{2}
\frac{\widetilde{\mathrm{g}}^{2}}{\omega_{\mathrm{g}}^{2}}
+
\hbar^{2}\widetilde{\mathrm{g}}^{2}
\sum_{\nu=1}^{\infty}
\frac{|\langle\widetilde{E}_{\nu}^{\mathrm{ren}}|E_{0}^{\mathrm{ren}}\rangle|^{2}}{
(E_{\nu}-E_{0}+\hbar\omega_{\mathrm{g}})^{2}
}
\label{eq:photon_number_expression}
\end{equation}
by using the pull-through formula. 
We note that we do not use the perturbation theory 
to prove this expression. 
For the operator-theoretical pull-through formula, 
see Ref.\cite{hiro-rims}.  
In the same way, for the ground-state expectation of 
bare photon, we have the expression, 
$$
N_{0}^{\mathrm{bare}}=
|\langle\widetilde{E}_{0}|E_{0}\rangle|^{2}
\frac{\mathrm{g}^{2}}{\omega_{\mathrm{c}}^{2}}
+
\hbar^{2}\mathrm{g}^{2}
\sum_{\nu=1}^{\infty}
\frac{|\langle\widetilde{E}_{\nu}|E_{0}\rangle|^{2}}{
(E_{\nu}-E_{0}+\hbar\omega_{\mathrm{c}})^{2}
}. 
$$
Eq.(\ref{eq:photon_number_expression}) implies the inequality, 
\begin{equation}
N_{0}^{\mathrm{ren}}\le 
|\langle\widetilde{E}_{0}^{\mathrm{ren}}|E_{0}^{\mathrm{ren}}\rangle|^{2}
\frac{\widetilde{\mathrm{g}}^{2}}{\omega_{\mathrm{g}}^{2}}
+
\hbar^{2}\widetilde{\mathrm{g}}^{2}
\sum_{\nu=1}^{\infty}
\frac{\hbar^{2}\omega_{\mathrm{a}}^{2}}{
(E_{\nu}-E_{0}+\hbar\omega_{\mathrm{g}})^{2}
(E_{\nu}-E_{0})^{2}
}. 
\label{eq:upp-bnd*}
\end{equation}
Here, we note the followings before we prove Eqs.(\ref{eq:photon_number_expression}) 
and (\ref{eq:upp-bnd*}). 
The set of $\{|\widetilde{E}^{\mathrm{ren}}_{\nu}\rangle\}_{\nu=0}^{\infty}$ 
is a complete orthonormal system of the Hilbert space 
$\sigma_{x}\mathcal{F}=\mathcal{F}$ 
by Theorem 4.3 of Ref.\cite{hiro-iumj}. 
Thus, there is an $\nu_{*}$ such that 
$\langle\widetilde{E}_{\nu_{*}}|E_{0}\rangle\ne 0$ 
since $|E_{0}\rangle\ne 0$. 
This fact gives a lower bound of the ground-state expectation 
of physical photon: 
\begin{equation}
\hbar^{2}\widetilde{\mathrm{g}}^{2}\, 
\frac{|\langle\widetilde E_{\nu_{*}}^{\mathrm{ren}}|
E_{0}^{\mathrm{ren}}\rangle|^{2}}{
(E_{\nu_{*}}-E_{0}+\hbar\omega_{\mathrm{g}})^{2}}
\le N_{0}^{\mathrm{ren}}. 
\label{eq:lower-bnd}
\end{equation} 
In the case $\varepsilon=0$, we have $\nu_{*}\ne 0$ 
since we can show $\langle\widetilde E_{0}^{\mathrm{ren}}|
E_{0}^{\mathrm{ren}}\rangle=0$ using the parity symmetry, 
$[ H(\omega_{\mathrm{a}}, 0, \omega_{\mathrm{g}},\widetilde{\mathrm{g}}) , 
\Pi]=0$, 
for the parity operator $\Pi$. 
Eqs.(\ref{eq:photon_number_expression}) 
and (\ref{eq:lower-bnd}) tell us that 
the ground state has at least 
a physical photon 
with the transition probability, 
$|\langle\widetilde E_{\nu_{*}}^{\mathrm{ren}}|
E_{0}^{\mathrm{ren}}\rangle|^{2}$, 
form the states $|\widetilde E_{\nu_{*}}^{\mathrm{ren}}\rangle$ 
on the other side of the mirror.

We now prove Eqs.(\ref{eq:photon_number_expression}) 
and (\ref{eq:upp-bnd*}): 
We first give the so-called pull-through formula for the quantum Rabi Hamiltonian. 
We note the equation, 
$$
[H(\omega_{\mathrm{a}},\varepsilon,\omega_{\mathrm{g}}, \widetilde{\mathrm{g}})
-E_{0}, a]|E_{0}^{\mathrm{ren}}\rangle
=(H(\omega_{\mathrm{a}},\varepsilon,\omega_{\mathrm{g}}, \widetilde{\mathrm{g}})
-E_{0})a|E_{0}^{\mathrm{ren}}\rangle.
$$
Meanwhile, we compute the commutator in the left hand side of the above as 
$$
[H(\omega_{\mathrm{a}},\varepsilon,\omega_{\mathrm{g}}, \widetilde{\mathrm{g}})-E_{0}, a]
=\, -\hbar\omega_{\mathrm{g}}a-\hbar\widetilde{\mathrm{g}}\sigma_{x}.
$$
Combining these equations, we obtain the pull-through formula: 
\begin{equation}
a|E_{0}^{\mathrm{ren}}\rangle
=\, -\hbar\widetilde{\mathrm{g}}
\left( H(\omega_{\mathrm{a}},\varepsilon,\omega_{\mathrm{g}}, \widetilde{\mathrm{g}})
-E_{0}+\hbar\omega_{\mathrm{g}}\right)^{-1}\sigma_{x}|E_{0}^{\mathrm{ren}}\rangle. 
\label{eq:pull-through}
\end{equation}
Using Eq.(\ref{eq:pull-through}), we have the expression of 
the ground-state expectation of photon as 
\begin{equation}
N_{0}^{\mathrm{ren}}=
\hbar^{2}\widetilde{\mathrm{g}}^{2}
\| \left( H(\omega_{\mathrm{a}},\varepsilon,\omega_{\mathrm{g}}, \widetilde{\mathrm{g}})
-E_{0}+\hbar\omega_{\mathrm{g}}\right)^{-1}\sigma_{x}|E_{0}^{\mathrm{ren}}\rangle
\|_{\mathcal{F}}^{2}, 
\label{eq:5-1}
\end{equation}
which implies the upper bound Eq.(\ref{eq:upp-bnd}). 
Using the completeness of the set 
$\left\{|E_{\nu}^{\mathrm{ren}}\rangle\right\}_{\nu=0}^{\infty}$, 
we have the expansion, 
\begin{equation}
\sigma_{x}|E_{0}^{\mathrm{ren}}\rangle
=\sum_{\nu=0}^{\infty}
\langle E_{\nu}^{\mathrm{ren}}|\sigma_{x}|E_{0}^{\mathrm{ren}}\rangle
|E_{\nu}^{\mathrm{ren}}\rangle.
\label{eq:5-2}
\end{equation}
Inserting Eq.(\ref{eq:5-2}) into Eq.(\ref{eq:5-1}), 
we reach Eq.(\ref{eq:photon_number_expression}). 

To obtain Eq.(\ref{eq:upp-bnd*}), 
we have only to estimate the transition probability, 
$|\langle\widetilde{E}_{\nu}^{\mathrm{ren}}|E_{0}^{\mathrm{ren}}\rangle|^{2}$. 
Define the spin-chiral transformed Hamiltonian 
$\widetilde{H}(\omega_{\mathrm{a}},\varepsilon,\omega_{\mathrm{g}},
\widetilde{\mathrm{g}})$ by 
$$
\widetilde{H}(\omega_{\mathrm{a}},\varepsilon,\omega_{\mathrm{g}},
\widetilde{\mathrm{g}})
:=\sigma_{x}H(\omega_{\mathrm{a}},\varepsilon,\omega_{\mathrm{g}},
\widetilde{\mathrm{g}})\sigma_{x}. 
$$  
Then, we have the equation,
\begin{equation}
\widetilde{H}(\omega_{\mathrm{a}},\varepsilon,\omega_{\mathrm{g}},
\widetilde{\mathrm{g}})
=H(-\omega_{\mathrm{a}},\varepsilon,\omega_{\mathrm{g}},
\widetilde{\mathrm{g}}). 
\label{eq:8-1}
\end{equation}
Since $|\widetilde{E}_{\nu}^{\mathrm{ren}}\rangle$ is an eigenstate of the 
transformed Hamiltonian $\widetilde{H}(\omega_{\mathrm{a}},\varepsilon,
\omega_{\mathrm{g}},\widetilde{\mathrm{g}})$ 
with the eigenenergy $E_{\nu}$, 
Eq.(\ref{eq:8-1}) leads to the following: 
\begin{eqnarray*}
E_{\nu}\langle\widetilde{E}_{\nu}^{\mathrm{ren}}|E_{0}^{\mathrm{ren}}\rangle
&=&
\langle\widetilde{E}_{\nu}^{\mathrm{ren}}|H(-\omega_{\mathrm{a}},\varepsilon,
\omega_{\mathrm{g}},\widetilde{\mathrm{g}})|E_{0}^{\mathrm{ren}}\rangle \\ 
&=&\langle\widetilde{E}_{\nu}^{\mathrm{ren}}|H(\omega_{\mathrm{a}},\varepsilon,
\omega_{\mathrm{g}},\widetilde{\mathrm{g}})
-\hbar\omega_{\mathrm{a}}\sigma_{z}|E_{0}^{\mathrm{ren}}\rangle \\ 
&=&E_{0}\langle\widetilde{E}_{\nu}^{\mathrm{ren}}|E_{0}^{\mathrm{ren}}\rangle
-\hbar\omega_{\mathrm{a}}\langle\widetilde{E}_{\nu}^{\mathrm{ren}}|
\sigma_{z}|E_{0}^{\mathrm{ren}}\rangle,
\end{eqnarray*}
which implies the equation, 
$(E_{\nu}-E_{0})\langle\widetilde{E}_{\nu}^{\mathrm{ren}}|E_{0}^{\mathrm{ren}}\rangle
=\, -\hbar\omega_{\mathrm{a}}
\langle\widetilde{E}_{\nu}^{\mathrm{ren}}|\sigma_{z}|E_{0}^{\mathrm{ren}}\rangle$, 
and thus, 
$$
%\begin{equation}
(E_{\nu}-E_{0})|\langle\widetilde{E}_{\nu}^{\mathrm{ren}}|E_{0}^{\mathrm{ren}}\rangle|
=\hbar\omega_{\mathrm{a}}
|\langle\widetilde{E}_{\nu}^{\mathrm{ren}}|\sigma_{z}|E_{0}^{\mathrm{ren}}\rangle|.
$$
Therefore, we can obtain our desired estimate:  
$$
|\langle\widetilde{E}_{\nu}^{\mathrm{ren}}|E_{0}^{\mathrm{ren}}\rangle|^{2}
\le \frac{\hbar^{2}\omega_{\mathrm{a}}^{2}}{
(E_{\nu}-E_{0})^{2}}. 
$$

We now prove Eq.(\ref{eq:lower-bnd-2}). 
By the variational principle, we have the inequality, 
\begin{eqnarray*}
E_{0}&=&\langle E_{0}^{\mathrm{ren}}|
H(\omega_{\mathrm{a}},\varepsilon,
\omega_{\mathrm{g}},\widetilde{\mathrm{g}})
|E_{0}^{\mathrm{ren}}\rangle \\  
&=&\langle E_{0}^{\mathrm{ren}}|H_{\mathrm{atom}}(\omega_{\mathrm{a}},
\varepsilon)|E_{0}^{\mathrm{ren}}\rangle 
+\langle E_{0}^{\mathrm{ren}}|H_{\mathrm{cavity}}(\omega_{\mathrm{g}})
|E_{0}^{\mathrm{ren}}\rangle \\ 
&{}&
+\langle E_{0}^{\mathrm{ren}}|H_{\mathrm{int}}(\mathrm{g})|E_{0}^{\mathrm{ren}}\rangle \\ 
&\ge& -\, \frac{\hbar}{2}\sqrt{\omega_{\mathrm{a}}^{2}+\varepsilon^{2}}
+\hbar\omega_{\mathrm{g}}N_{0}^{\mathrm{ren}}+\frac{\hbar\omega_{\mathrm{g}}}{2}
+\hbar\widetilde{\mathrm{g}}
\langle E_{0}^{\mathrm{ren}}|\sigma_{x}(a+a^{\dagger})|E_{0}^{\mathrm{ren}}\rangle \\ 
&\ge& 
 -\, \frac{\hbar}{2}\sqrt{\omega_{\mathrm{a}}^{2}+\varepsilon^{2}}
+\hbar\omega_{\mathrm{g}}N_{0}^{\mathrm{ren}}+\frac{\hbar\omega_{\mathrm{g}}}{2}
-\hbar\widetilde{\mathrm{g}}
|\langle E_{0}^{\mathrm{ren}}|\sigma_{x}(a+a^{\dagger})|E_{0}^{\mathrm{ren}}\rangle|. 
\end{eqnarray*}
Using the Schwarz inequality, the fact that the creation operator $a^{\dagger}$ is the adjoint 
operator of the annihilation operator $a$, and Eq.(3.21) of Ref.\cite{hiro-iumj}, 
we have the inequality, 
$$
|\langle E_{0}^{\mathrm{ren}}|\sigma_{x}(a+a^{\dagger})|E_{0}^{\mathrm{ren}}\rangle|
\le 2\| a|E_{0}^{\mathrm{ren}}\rangle\|_{\mathcal{F}} \\ 
=2\sqrt{N_{0}^{\mathrm{ren}}}.
$$
Combining above two inequalities, together with Eq.(\ref{eq:gse-1}), 
leads to the inequality, 
$$
2\hbar\widetilde{\mathrm{g}}\sqrt{N_{0}^{\mathrm{ren}}}
-\hbar\omega_{\mathrm{g}}N_{0}^{\mathrm{ren}}
\ge \hbar F(\mathrm{g}), 
$$
where 
$$
F^{\mathrm{ren}}(\mathrm{g})
=\frac{1}{2}\sqrt{\omega_{\mathrm{a}}^{2}+\varepsilon^{2}}
\left( e^{-2\widetilde{g}^{2}/\omega_{\mathrm{g}}^{2}}-1\right)
+\frac{\widetilde{\mathrm{g}}^{2}}{\omega_{\mathrm{g}}}.
$$
Set $X$ and $\gamma$ as $X:=\sqrt{\omega_{\mathrm{g}}N_{0}^{\mathrm{ren}}}$ and 
$\gamma:=\sqrt{\widetilde{\mathrm{g}}^{2}/\omega_{\mathrm{g}}}$. 
Then, dividing the both sides by $\hbar$, 
the above inequality is rewritten as 
$2\gamma X-X^{2}\ge F^{\mathrm{ren}}(\mathrm{g})$. 
Thus, $X$ has to satisfy 
$\gamma-\sqrt{\gamma^{2}-F^{\mathrm{ren}}(\mathrm{g})}
\le X\le \gamma+\sqrt{\gamma^{2}-F^{\mathrm{ren}}(\mathrm{g})}$, 
which implies 
$$
\sqrt{\frac{\widetilde{\mathrm{g}}^{2}}{\omega_{\mathrm{g}}}}
-\sqrt{\frac{\widetilde{\mathrm{g}}^{2}}{\omega_{\mathrm{g}}}-F^{\mathrm{ren}}(\mathrm{g})}
\le \sqrt{\omega_{\mathrm{g}}N_{0}}.  
$$
Thus, we obtain our desired statement.

\subsection{Possibility of Entangled Ground State}

We consider the possibility 
that the ground state becomes an entangled state. 

Eq.(\ref{eq:lower-bnd}) says that once the coupling strength 
is turned on (i.e., $\mathrm{g}>0$), 
the renormalized ground-state expectation of the physical photon 
is strictly positive (i.e., $N_{0}^{\mathrm{ren}}>0$). 
Thus, there are a spin state $|s_{*}\rangle$ and the Fock state $|n_{*}\rangle$ 
with a photon number $n_{*}>0$ so that  
\begin{equation}
\langle n_{*}, s_{*}|E_{0}^{\mathrm{ren}}\rangle\ne 0.
\label{eq:4-16-1}
\end{equation}
More precisely, let us express $|E_{0}^{\mathrm{ren}}\rangle$ as 
$|E_{0}^{\mathrm{ren}}\rangle
=\sum_{n=0}^{\infty}
c_{n}^{\uparrow}|\uparrow,n\rangle
+\sum_{n=0}^{\infty}
c_{n}^{\downarrow}|\downarrow,n\rangle$. 
We have the expression of the renormalized ground-state expectation as 
$$
N_{0}^{\mathrm{ren}}
\equiv 
\langle E_{0}^{\mathrm{ren}}|
a^{\dagger}a|E_{0}^{\mathrm{ren}}\rangle
=\sum_{n=1}^{\infty}
n|c_{n}^{\uparrow}|^{2}
+\sum_{n=1}^{\infty}
n|c_{n}^{\downarrow}|^{2}.
$$
Since we know $N_{0}^{\mathrm{ren}}>0$, 
we eventually obtain that there are $s_{*}\in\{\downarrow,\uparrow\}$ 
and a natural number $n_{*}>0$ so that 
$|\langle n_{*}, s_{*}|E_{0}^{\mathrm{ren}}\rangle|^{2}
=|c_{n_{*}}^{s_{*}}|^{2}\ne 0$. 
That is, we reach Eq.(\ref{eq:4-16-1}).

We immediately realize that the quantum state $|s_{*}, n_{*}\rangle$ 
is not a ground state, and therefore, there is another quantum state 
$|s, n\rangle$ so that $|\langle n, s|E_{0}^{\mathrm{ren}}\rangle|^{2}\ne 0$. 
We try to seek it now. 
We define the orthogonal projection operators, 
$P_{|\sharp\rangle}$ and $P_{|0\rangle}$, 
for $\sharp=\uparrow, \downarrow$ by 
$P_{|\sharp\rangle}:=\langle\sharp|\,\,\,\rangle|\sharp\rangle$ 
and 
$P_{|0\rangle}:=\langle 0|\,\,\,\rangle|0\rangle$. 
We have the inequality, 
$P_{|0\rangle}\ge 1-a^{\dagger}a$, 
as shown in the proof of Lemma 4.6 
of Ref.\cite{arai-hirokawa} 
since 
$a^{\dagger}a=\sum_{n=1}^{\infty}nP_{|n\rangle}$ 
and $1=\sum_{n=0}^{\infty}P_{|n\rangle}$ for the orthogonal 
projection defined by $P_{|n\rangle\rangle}:=
\langle n|\,\,\,\rangle|n\rangle$, 
we have 
\begin{eqnarray*}
|\langle 0, \uparrow|E_{0}^{\mathrm{ren}}\rangle|^{2}
+|\langle 0, \downarrow|E_{0}^{\mathrm{ren}}\rangle|^{2}
&=&
\langle E_{0}^{\mathrm{ren}}|P_{|\uparrow\rangle}\otimes P_{|0\rangle}|
E_{0}^{\mathrm{ren}}\rangle
+\langle E_{0}^{\mathrm{ren}}|P_{|\downarrow\rangle}\otimes P_{|0\rangle}|
E_{0}^{\mathrm{ren}}\rangle 
\nonumber \\ 
&=& \langle E_{0}^{\mathrm{ren}}|I\otimes P_{|0\rangle}|E_{0}^{\mathrm{ren}}\rangle 
\nonumber \\ 
&\ge& \langle E_{0}^{\mathrm{ren}}|1-a^{\dagger}a|E_{0}^{\mathrm{ren}}\rangle 
=1-N_{0}^{\mathrm{ren}}. 
\end{eqnarray*}
By this inequality, we can say that there is a spin-state $|s_{0}\rangle$ so that  
\begin{equation}
\langle 0, s_{0}|E_{0}^{\mathrm{ren}}\rangle\ne 0\quad 
\textrm{if}\quad N_{0}^{\mathrm{ren}}<1. 
\label{eq:4-16-2}
\end{equation}
Eqs.(\ref{eq:4-16-1}) and (\ref{eq:4-16-2}) tell us that 
the expansion of the ground state $|E_{0}^{\mathrm{ren}}\rangle$ by 
the states $|\sharp, n\rangle$ must include a superposition of 
the state $|s_{0}, 0\rangle$ and $|s_{*}, n_{*}\rangle$ 
when the ground-state expectation of photon satisfies 
$0<N_{0}^{\mathrm{ren}}<1$. 
Therefore, at least a quantum states, 
$|s_{0}, 0\rangle$ and $|s_{*}, n_{*}\rangle$, 
make an entanglement in the expansion of 
the ground state $|E_{0}^{\mathrm{ren}}\rangle$ then.  
As for the possibility of the entangled ground state, 
see the following section.

\section{Summary and Discussion} 

For the generalized quantum Rabi model with the $A^{2}$-term, 
we gave an upper bound and a lower bound 
of the renormalized ground-state expectation of physical photon. 
We showed how the ground state has at least one physical photon, 
and the possibility of entanglement in the ground state. 
As shown in Eq.(\ref{eq:approximation_coherent_state}) 
for the quantum Rabi model, 
its ground state is well approximated by the coherent state 
for sufficiently large coupling strength. 
Therefore, we conjecture that both the bare and physical 
ground states makes a highly entangled ground state. 
We will show this conjecture and its leading term. 
Our results say that we may stably store a physical photon 
in the ground state for a properly large coupling strength at least. 
Quoted from Ciuti's lecture \cite{ciuti2}: The ground state is the lowest energy state; 
the photons in the ground state cannot escape the cavity.

\appendix 

\section{Theorem for Improvement of Convergence}
\label{app:theorem}

In this appendix, we state and prove the theorem that we used in \S\ref{sec:hlp}. 

\textsc{Theorem}: 
\textit{Let $A$ and $B$ be compact operators on a Hilbert space} $\mathfrak{H}$. 
\textit{If the operator sequence $W_{n}$ is bounded (i.e., 
$\sup_{n}\| W_{n}\|_{\mathrm{op}}<\infty$), 
and converges to $0$ as $n\to\infty$ 
in the weak operator topology, 
then the operator $AW_{n}B$ converges to $0$ in the uniform operator topology, 
i.e., in the operator norm.}

\textsc{Proof}: Let $\left\{\varphi_{j}\right\}_{j=1}^{\infty}$ be a complete orthonormal 
system of the Hilbert space $\mathfrak{H}$. By Theorem VI.13 of Ref.\cite{rs1}, 
the finite rank operator $B_{N}:=\sum_{j=1}^{N}\left(\varphi_{j},\cdot\right)_{\mathfrak{H}}
B\varphi_{j}$ converges to the compact operator $B$ as $N\to\infty$ in the uniform operator topology. 
By the triangle inequality, we have 
\begin{eqnarray}
\|AW_{n}B\|_{\mathrm{op}}
&\le& \|AW_{n}B_{N}\|_{\mathrm{op}}+\|AW_{n}(B-B_{N})\|_{\mathrm{op}} 
\nonumber \\ 
&\le& \|AW_{n}B_{N}\|_{\mathrm{op}}+
\| A\|_{\mathrm{op}}(\sup_{n}\|W_{n}\|_{\mathrm{op}})\|B-B_{N}\|_{\mathrm{op}} 
\label{eq:th-0}
\end{eqnarray}
for arbitrary natural number $N$. 
We can estimate the first term of the above as
\begin{eqnarray*}
\|AW_{n}B_{N}\|_{\mathrm{op}}
&=&\sup_{\|\psi\|_{\mathfrak{H}}=1}
\|AW_{n}B_{N}\psi\|_{\mathfrak{H}}
=
\sup_{\|\psi\|_{\mathfrak{H}}=1}
\Biggl\|
\sum_{j=1}^{N}\left(\varphi_{j},\psi\right)_{\mathfrak{H}}
AW_{n}B\varphi_{j}
\Biggr\|_{\mathfrak{H}} \\ 
&\le& 
\sum_{j=1}^{N}\| AW_{n}B\varphi_{j}\|_{\mathfrak{H}}, 
\end{eqnarray*}
where we used the Schwarz inequality, 
$|\left(\varphi_{j},\psi\right)_{\mathfrak{H}}|\le 
\|\varphi_{j}\|_{\mathfrak{H}}\|\psi\|_{\mathfrak{H}}=1$. 
Since the operators $W_{n}$ converges to $0$ in the weak operator topology, 
the vectors $\left\{ W_{n}B\varphi_{j}\right\}_{n}$ 
are weakly convergent sequence. 
Since the operator $A$ is compact, 
the vectors $\left\{ AW_{n}B\varphi_{j}\right\}_{n}$ 
are a convergent sequence in the Hilbert space $\mathfrak{H}$ 
(i.e., the sequence $\left\{ AW_{n}B\varphi_{j}\right\}_{n}$ converges 
in the norm of the Hilbert space $\mathfrak{H}$) 
by Theorem VI.11 of Ref.\cite{rs1}. 
Thus, we have 
\begin{equation}
\lim_{n\to\infty}\|AW_{n}B_{N}\|_{\mathrm{op}}=0\,\,\, 
\textrm{for arbitrary natural number $N$}.
\label{eq:th-1}
\end{equation}
Combining Eqs.(\ref{eq:th-0}) and (\ref{eq:th-1}) leads to 
the inequality, 
$$
\lim_{n\to\infty}
\|AW_{n}B\|_{\mathrm{op}}
\le \| A\|_{\mathrm{op}}(\sup_{n}\|W_{n}\|_{\mathrm{op}})\|B-B_{N}\|_{\mathrm{op}} 
$$
for arbitrary natural number $N$. 
Taking the limit $N\to\infty$ in the above, we conclude our theorem. 
\qquad $\square$

\section{Estimate of $N_{0}^{\mathrm{bare}}$}
\label{app:bare}

We easily have 
\begin{equation}
(\Delta\Phi)^{2}\le \frac{1}{2\omega_{\mathrm{c}}}\langle E_{0}|
(a+a^{\dagger})^{2}|E_{0}\rangle
\label{eq:5-15-1}
\end{equation}
since $(\Delta\Phi)^{2}=\langle E_{0}|\Phi^{2}|E_{0}\rangle
-\langle E_{0}|\Phi|E_{0}\rangle^{2}$, 
and 
\begin{equation}
\langle E_{0}|
(a+a^{\dagger})^{2}|E_{0}\rangle
=2N_{0}^{\mathrm{bare}}+1
+\langle E_{0}|
a^{2}+(a^{\dagger})^{2}|E_{0}\rangle. 
\label{eq:5-15-2}
\end{equation}
Meanwhile, 
by Eqs.(3.21) and (3.22) of Ref.\cite{hiro-iumj}, 
we have 
$\| a|E_{0}\rangle\|_{\mathcal{F}}=\sqrt{N_{0}^{\mathrm{bare}}}$ 
and 
$\| a^{\dagger}|E_{0}\rangle\|_{\mathcal{F}}=\sqrt{N_{0}^{\mathrm{bare}}+1}$.  
With the help of the Schwarz inequality, 
these equations tell us  
\begin{equation}
|\langle E_{0}|(a^{\sharp})^{2}|E_{0}\rangle|
\le 
\| a|E_{0}\rangle\|_{\mathcal{F}}
\| a^{\dagger}|E_{0}\rangle\|_{\mathcal{F}}
\le \sqrt{N_{0}^{\mathrm{bare}}}
\sqrt{N_{0}^{\mathrm{bare}}+1}
\le 
\frac{2N_{0}^{\mathrm{bare}}+1}{2}.
\label{eq:5-15-3}
\end{equation}
Combining Eqs.(\ref{eq:5-15-1})--(\ref{eq:5-15-3}) 
leads to Eq.(\ref{eq:field_fluctuation-bare}). 

Since $b=UaU^{*}$ and $b^{\dagger}=Ua^{\dagger}U^{*}$, 
we have the relations
$$
U^{*}aU=M_{1}a+M_{2}a^{\dagger}
\quad\textrm{and}\quad 
U^{*}a^{\dagger}U=M_{2}a+M_{1}a^{\dagger}
%\label{eq:result0'}
$$
by Eq.(\ref{eq:result0}). 
Using these relations, 
we can rewrite the ground-state expectation of bare photon as 
\begin{eqnarray}
N_{0}^{\mathrm{bare}}&=&
\langle E_{0}^{\mathrm{ren}}|(U^{*}a^{\dagger}U)(U^{*}aU)|E_{0}^{\mathrm{ren}}\rangle 
\nonumber \\ 
&=& 
(M_{1}^{2}+M_{2}^{2})N_{0}^{\mathrm{ren}}+M_{2}^{2}
+M_{1}M_{2}\langle E_{0}^{\mathrm{ren}}|
a^{2}+(a^{\dagger})^{2}|E_{0}^{\mathrm{ren}}\rangle. 
\label{eq:5-15-4}
\end{eqnarray}
In the same way as showing Eq.(\ref{eq:5-15-3}), 
we have the inequality, 
$|\langle E_{0}^{\mathrm{ren}}|
a^{2}+(a^{\dagger})^{2}|E_{0}^{\mathrm{ren}}\rangle|
\le 2\sqrt{N_{0}^{\mathrm{ren}}}\sqrt{N_{0}^{\mathrm{ren}}+1}$. 
Since $M_{1}>0$ and $M_{2}<0$, 
we obtain the estimate from below by Eqs.(\ref{eq:upp-bnd}), (\ref{eq:5-15-4}), and (\ref{eq:5-15-5}): 
\begin{eqnarray}
N_{0}^{\mathrm{bare}}&\ge& 
(M_{1}^{2}+M_{2}^{2})N_{0}^{\mathrm{ren}}
+M_{2}^{2}
+2M_{1}M_{2}\sqrt{N_{0}^{\mathrm{ren}}}\sqrt{N_{0}^{\mathrm{ren}}+1} \nonumber \\ 
&\ge& 
M_{2}^{2}
+2M_{1}M_{2}(N_{0}^{\mathrm{ren}}+1) 
\nonumber \\ 
&\ge& 
M_{2}^{2}
-\epsilon(\mathrm{g}),
\label{eq:5-15-5}
\end{eqnarray}
where 
\begin{eqnarray*}
\epsilon(\mathrm{g})
&=&
\frac{1}{2}\left\{
\left(\frac{\omega_{\mathrm{g}}^{2}-\omega_{\mathrm{c}}^{2}}{
\omega_{\mathrm{c}}\omega_{\mathrm{g}}}\right)
\left(
\frac{\widetilde{g}^{2}}{\omega_{\mathrm{g}}}+1
\right)\right\} \\ 
&=& 
\frac{1}{2}\Biggl\{
\left(\frac{(\omega_{\mathrm{g}}/\mathrm{g}^{(\ell+1)/2})^{2}
-(\omega_{\mathrm{c}}/\mathrm{g}^{(\ell+1)/2})^{2}}{
\omega_{\mathrm{c}}(\omega_{\mathrm{g}}/\mathrm{g}^{(\ell+1)/2})}\right) 
%\\ 
%&{}&\qquad\qquad
\left(
\frac{\omega_{\mathrm{c}}}{(\omega_{\mathrm{g}}/\mathrm{g}^{(\ell+1)/2})^{3}}
\mathrm{g}^{1-\ell}
+
\frac{1}{\mathrm{g}^{(\ell+1)/2}}
\right)\Biggr\}.
\end{eqnarray*}
Therefore, we finally obtain our desired lower estimate 
from Eq.(\ref{eq:5-15-5}).

\section{Numerical-Analysis Method for Eigenvalues of Rabi Hamiltonian}
\label{app:nam}

We recall that in the case $\varepsilon=0$, the quantum Rabi Hamiltonian 
$H_{\mathrm{Rabi}}(\omega_{\mathrm{a}},\omega_{\mathrm{c}},\mathrm{g}):=
H(\omega_{\mathrm{a}}, 0, \omega_{\mathrm{c}}, \mathrm{g})$ 
has the parity symmetry: 
$[ H_{\mathrm{Rabi}}(\omega_{\mathrm{a}},\omega_{\mathrm{c}},\mathrm{g}), 
\Pi]=0$ for the parity operator $\Pi=\sigma_{z}(-1)^{a^{\dagger}a}$. 
Let us denote by $\mathcal{F}_{\pm 1}$ the eigenspace of the parity operator 
with the eigenvalue, $\pm 1$. 
We can decompose the state space $\mathcal{F}$ as 
$\mathcal {F}=\mathcal{F}_{-1}\oplus\mathcal{F}_{+1}$, 
and then, we can also decompose the quantum Rabi Hamiltonian into the direct sum of the two self-adjoint 
operators $H^{\pm}$ acting in the individual state space $\mathcal{F}_{\pm 1}$ 
in the same way as in Eq.(\ref{eq:parity-decomposition}), 
\begin{equation}
H_{\mathrm{Rabi}}(\omega_{\mathrm{a}},\omega_{\mathrm{c}},\mathrm{g})
=H^{-}\oplus H^{+},
\label{eq:parity_decomposition_Rabi-Hamiltonian}
\end{equation}
where $H^{\pm}$ is obtained by subtracting the $A^{2}$-term from 
$H_{\pm}$ in Eq.(\ref{eq:parity-decomposition2}). 
Following this decomposition, 
in the numerical analysis for the eigenvalue problem for the Rabi Hamiltonian 
$H_{\mathrm{Rabi}}(\omega_{\mathrm{a}},\omega_{\mathrm{c}},\mathrm{g})$, 
we have only to seek the eigenvalues of $H^{\pm}$.  
We can give the matrix-representation $H^{\pm}_{\mathrm{matrix}}$ 
of the operator $H^{\pm}$ in the following. 
Using the complete 
orthonormal basis, 
$|\!\!\uparrow, 0\rangle$, $|\!\!\downarrow, 1\rangle$, 
$|\!\!\uparrow, 2\rangle$, $|\!\!\downarrow, 3\rangle$, 
$\cdots$, for $H^{+}$, 
and using the complete orthonormal basis, 
$|\!\!\downarrow, 0\rangle$, $|\!\!\uparrow, 1\rangle$, 
$|\!\!\downarrow, 2\rangle$, $|\!\!\uparrow, 3\rangle$, 
$\cdots$, for $H^{-}$, they are 
\begin{equation}
H^{\pm}_{\mathrm{matrix}}:=
\left(
\begin{array}{ccccc}
     d_{0}\pm\Delta   & \sqrt{1}g & \quad       & \quad      & \quad \\
     \sqrt{1}g & d_{1}\mp\Delta  & \sqrt{2}g &  \quad     & \quad \\
     \quad       & \sqrt{2}g & d_{2}\pm\Delta   & \sqrt{3}g & \quad \\
     \quad       & \quad       & \sqrt{3}g & d_{3}\mp\Delta  & \ddots \\
     \quad       & \quad       & \quad       & \ddots      & \ddots \\
      \quad      & \quad       & \quad       & \quad       & \quad 
\end{array}
\right),  
\label{eq:matrix_rep}
\end{equation}
where $\Delta=\hbar\omega_{\mathrm{a}}/2$, $g=\hbar\mathrm{g}$, 
and $d_{n}=\hbar\omega_{\mathrm{c}}(n+2^{-1})$. 
With these matrix representations 
we can seek the eigenvalues of the operators $H^{\pm}$. 
To achieve this, we make good use of the eigenvalue problem 
for the symmetric Jacobi matrix: 
For given complex sequences, $a = \{a_{n}\}_{n=0}^{\infty}$ and 
$b = \{b_{n}\}_{n=0}^{\infty}$, the matrix 
$$
J(a,b):=
\left(
\begin{array}{cccccc}
 a_{0}            & b_{0}            & \quad           
     & \quad           & \quad   & \quad    \\
 \bar{b}_{0} & a_{1}            & b_{1}            
     & \quad           & \quad     & \quad   \\
 \quad            & \bar{b}_{1} & a_{2}            
     & b_{2}           &  \quad     & \quad  \\
 \quad            & \quad            & \bar{b}_{2} 
     & a_{3}           & b_{3}     & \quad \\
 \quad            & \quad            & \quad            
     & \bar{b}_{3} & a_{4}     & \ddots  \\
 \quad            & \quad            & \quad             
     & \quad           & \ddots    & \ddots
\end{array}
\right)
$$ 
is called the symmetric Jacobi matrix. 
We assume the following two conditions:
\begin{enumerate}
 \item[(A.1)] $a_n \to \infty$ as $n\to\infty$.
 \item[(A.2)] $(|b_n|+|b_{n-1}|) (|a_n|+1)^{-1} \to 0$ as $n\to\infty$. 
\end{enumerate}
Thanks to these assumption, the operator consisting of 
the off-diagonal entries is the infinitesimally small 
with respect to the operator consisting of the diagonal entries, 
the Kato-Rellich theorem \cite{rs2} guarantees the operator $J(a,b)$ to be self-adjoint 
acting in the Hilbert space of the square-summable sequences. 
For any natural number $N$ we define a matrix $J_{N}(a,b)$ by 
$$
   J_N(a,b) := J(a,b) \Big|_{b_{N}=0}
 = \left(
    \begin{array}{cccc|ccc}
         a_0 & b_0    &               &         &               \\
   \bar{b}_0 & \ddots & \ddots        &         &               \\
             & \ddots & \ddots        & b_{N-1} &               \\
             &        & \bar{b}_{N-1} & a_N     & 0             &     \\  \hline
             &        &               & 0       & a_{N+1}       & b_{N+1} & \\
             &        &               &         & \bar{b}_{N+1} & \ddots & \ddots\\
             &        &               &         &               & \ddots & \ddots
    \end{array}
     \right).
     \qquad\qquad  
$$
Defining matrices $K_{N}(a,b)$ and $R_{N}(a,b)$ by 
$$
 K_{N}(a,b) = \left(
    \begin{array}{cccc}
     a_{0}  & b_{0}    &               &           \\
\bar{b}_{0} & \ddots & \ddots        &           \\
          & \ddots & a_{2}           & b_{N-1}  \\
          &        & \bar{b}_{N-1} & a_{N}     
     \end{array}
       \right)
       \,\,\,\textrm{and}\,\,\,
 R_N(a,b) = \left(
    \begin{array}{cccc}
     a_{N+1}    & b_{N+1}      &          &     \\
  \bar{b}_{N+1} & a_{N+2}      & b_{N+2}  &     \\
                & \bar{b}_{N+1}& a_{N+3}  &    \ddots \\
                &              & \ddots   &   \ddots
   \end{array}
     \right),
$$
we can decompose 
the matrix $J_{N}(a,b)$ into the direct sum of these matrices: 
\begin{equation}
J_{N}(a,b)=K_{N}(a,b) \oplus R_{N}(a,b).
\label{eq:4-16-3}
\end{equation}
Under the conditions (A.1) and (A.2), 
we will prove that the matrix $J_{N}(a,b)$ converges to 
the symmetric Jacob matrix $J(a,b)$ in the norm resolvent sense: 
\begin{equation}
\lim_{N\to\infty} \|(J_{N}(a,b)+i)^{-1} - (J(a,b)+i)^{-1}\| = 0, 
\label{eq:convergence_Jacobi}
\end{equation}
and that the $n$-th eigenvalue $\mu_{n}(K_N(a,b))$ of the matrix $K_{N}(a,b) $ 
converges to the $n$-th eigenvalue $\mu_{n}(J(a,b))$ of 
the symmetric Jacob matrix $J(a,b)$:
\begin{equation}
\mu_{n}(J(a,b)) = \lim_{N\to\infty}\mu_{n}(K_N(a,b)).
\label{eq:convergence_eigenvalue}
\end{equation}

We prove Eqs.(\ref{eq:convergence_Jacobi}) and 
(\ref{eq:convergence_eigenvalue}) from now on. 
We set $J$, $J_{N}$, $T$, $K_{N}$, and $R_{N}$ 
as $J:=J(a,b)$, $J_{N}:=J_{N}(a,b)$, 
$T:=J(a,0)$, $K_{N}:=K_{N}(a,b)$, and $R_{N}:=R_{N}(a,b)$ 
for simplicity. 
By the $2$nd resolvent equation we have 
\begin{eqnarray}
\|(J_{N}+i)^{-1}-(J+i)^{-1}\| 
&\le& \|(J-J_{N})(J+i)^{-1}\| 
\nonumber \\ 
&\le& \|(J-J_N)(T+i)^{-1}\|\, \|(T+i)(J+i)^{-1}\|. 
\label{eq:4-17-1}
\end{eqnarray} 
We note that the assumptions (A.1) and (A.2) guarantees 
$\|(T+i)(J+i)^{-1}\|<\infty$ 
in the right hand side of the inequality Eq.(\ref{eq:4-17-1}). 
Meanwhile, since $J-J_{N}
= b_{N}(|N+1\rangle\langle N|+|N\rangle\langle N+1|)$ 
by the definition, we reach the following inequality and the limit 
with the help of the condition (A.2): 
\begin{eqnarray} 
\|(J-J_N)(T+i)^{-1}\|
&\le& |b_{N}| \Big(\||N+1\rangle\langle N|(T+i)^{-1}\|
+\| |N\rangle\langle N+1|(T+i)^{-1}\|\Big) 
\nonumber \\
&=&|b_{N}|(|a_{N}+i|^{-1} + |a_{N+1}+i|^{-1})
\longrightarrow 0\,\,\, 
\textrm{as $N\to\infty$}. 
\label{eq:4-17-2}
\end{eqnarray} 
Eqs.(\ref{eq:4-17-1}) and (\ref{eq:4-17-2}) lead to 
Eq.(\ref{eq:convergence_Jacobi}). 

Thanks to Theorem VIII.23 of Ref.\cite{rs1}, we have 
the convergence, 
\begin{equation}
\lim_{N\to\infty} \mu_{n}(J_{N}) = \mu_{n}(J). 
\label{eq:4-17-3}
\end{equation}
Denote by $D_{N}$ 
the diagonal matrix of which diagonal entries are 
$a_{N+1}-|b_{N+1}|, a_{N+2}-|b_{N+1}|-|b_{N+2}|, 
a_{N+3}-|b_{N+2}|-|b_{N+3}|, a_{N+4}-|b_{N+3}|-|b_{N+4}|, \cdots$. 
Then, by the condition (A.2) we have the inequality and the limit,
$$
R_{N}
\ge D_{N}  
\ge 
\min_{n\ge N+1}(a_{n}-|b_{n-1}|-|b_{n}|)
\longrightarrow\infty\,\,\,\textrm{as ${N\to\infty}$}.  
$$
Thus, with this limit the mini-max principle (see Theorem XIII.1 of Ref.\cite{rs4}) 
says that 
\begin{equation}
\lim_{N\to\infty}\mu_n(R_N)=\infty. 
\label{eq:4-17-4}
\end{equation}
Remember that an eigenvalue of the matrix $J_{N}$ is 
an eigenvalue of the matrix $K_{N}$ or an eigenvalue of the matrix $R_{N}$ 
due to Eq.(\ref{eq:4-16-3}). 
Thus, Eq.(\ref{eq:4-17-4}) says that for sufficiently large $N$ we have 
\begin{equation}
\mu_{n}(J_{N}) = \mu_{n}(K_{N}). 
\label{eq:4-17-5}
\end{equation}
Eqs.(\ref{eq:4-17-3}) and (\ref{eq:4-17-5}) lead 
Eq.(\ref{eq:convergence_eigenvalue}).

\qquad 

\hfill\break 
\textbf{Acknowledgments}

\hfill\break 
M. H. thanks Kouichi Semba, Tomoko Fuse, and Fumiki Yoshihara 
for the valuable discussions on their experimental result, 
and Daniel Braak, I\~{n}iago Egusquiza, Elliott Lieb, 
and Kae Nemoto for their useful comments. 
He also  acknowledge J. S. M.'s hospitality at Aarhus University, 
and Franco Nori's at RIKEN. 
He also acknowledges the support from JSPS 
Grant-in-Aid for Scientific Researches 
(B) 26310210 and (C) 26400117. 
M. H and J. S. M. acknowledges the support from 
the International Network Program of 
the Danish Agency for Science, Technology and Innovation.

\qquad


\begin{thebibliography}{11}
\bibitem{nori}
Ashhab~S and Nori~F 2010 {\it Phys. Rev. A} \textbf{81} 042311
\bibitem{arai-hirokawa}
Arai~A and Hirokawa~M 1997 {\it  J. Funct. Anal.} {\bf 151} 455 
\bibitem{BCS}
Benenti~G, Casati~G, and Strini~G 2005 
{\it Principles of Quantum Computation and Information. Vol.I: Basic Concepts} 
(Singapore: World Scientific)
%\bibitem{berezin}
%Berezin~F~A 1966 {\it The Method of Second Quantization} (New York: Academic Press) 
\bibitem{braak1}
Braak D 2011 {\it Phys. Rev. Lett.} {\bf 107} 100401 
\bibitem{br}
Brown~L~M and Rechenberg~H 1996 
{\it The Origin of the Concept of Nuclear Forces}
(Bristol: IOP Publishing)
\bibitem{casanoba-solano}
Casanova~J, Romero~G, Lizuain~I, Gar\'{c}ia-Ripoll~J~J, 
and Solano~E 2010 
{\it Phys. Rev. Lett.} {\bf 105} 263603 
\bibitem{ciuti2}
Ciuti~C 2010 
{\it Ultrastrong coupling circuit QED:
vacuum degeneracy and quantum phase transitions} 
in the lecture at Coll\`{e}ge de France on June 8, 2010. 
\bibitem{enz}
Enz~C~P 1997 {\it Helv. Phys. Acta} {\bf 70} 141
\bibitem{henley-thirring}
Henley~E and Thirring~W 1962 
{\it Elementary Quantum Field Theory} 
(New York: McGraw-Hill)
\bibitem{hepp-lieb1}
Hepp~K and Lieb~E~H, 1973 {\it Ann. Phys.} {\bf 76} 360 
\bibitem{hepp-lieb2}
Hepp~K and Lieb~E~H, 1973 {\it Phys. Rev. A} {\bf 8} 2517 
\bibitem{hiro-rmp}
Hirokawa~M 2001 {\it Rev. Math. Phys.} {\bf 13} 221 
\bibitem{hiro-pla} 
Hirokawa~M 2002 
{\it Phys Lett. A} {\bf 294} 13 
\bibitem{hiro-rims}
Hirokawa~M 2006 {\it Pub. RIMS} {\bf 42} 897
\bibitem{hiro-pra}
Hirokawa~M 2009
{\it Phys. Rev. A} {\bf 79} 043408
\bibitem{hiro-iumj}
Hirokawa~M 2009 {\it Indiana Univ. Math. J.} {\bf 58} 1493
\bibitem{HHL} 
Hirokawa~M, Hiroshima~F and L\"{o}rinczi~J 2014 {\it Math. Z.} {\bf 277} 1165
\bibitem{hiro-hiro}
Hirokawa~M and Hiroshima~F 2014 {\it Comm. Stoch. Anal.} {\bf 8} 551 
\bibitem{hiro-qs}
Hirokawa~M 2015 {\it Quantum Studies: Math. Found.} {\bf 2} 379
\bibitem{hopfield}
Hopfield~J~J 1958 {\it Phys. Rev.} {\bf 112} 1555
%%% \bibitem{kato}
%%% Kato T 1995 {\it Perturbation Theory for Linear Operators} (Berlin: Springer Verlag) 
\bibitem{ciuti}
Nataf~P and Ciuti~C 2010 {\it Nature Comm.} \textbf{1} 72
\bibitem{preparata}
Preparata~G 1995 {\it QED Coherence in Matter} (Singapore: World Scientific)
\bibitem{rs1}
Reed~M and Simon~B 1980 {\it Method of Modern Mathematical Physics I. Functional Analysis} (San Diego: Academic Press) 
\bibitem{rs2}
Reed~M and Simon~B 1975 {\it Method of Modern Mathematical Physics  II. Fourier Analysis, 
Self-Adjointness} (San Diego: Academic Press) 
\bibitem{rs4}
Reed~M and Simon~B 1978 {\it Method of Modern Mathematical Physics  IV. Analysis of 
Operators} (San Diego: Academic Press) 
\bibitem{witten}
Witten~E 1981 {\it Nuclear Physics} B {\bf 185}513
\bibitem{solano-braak}
Wolf~F~A, Vallone~F, Romero~G, Kollar~M, 
Solano~E and Braak~D 2013  
{\it Phys. Rev. A} {\bf 87} 023835
\bibitem{semba2016}
Yoshinara~F, Fuse~T, Ashhab~S, Kakuyanagi~K, Saito~S and Semba~K, 
to appear in {\it Nature Phys.}, doi:10.1038/nphys3906, 
Published online 10 October 2016 
\end{thebibliography}
\end{document}